\documentclass{aastex}
\received{ }
\revised{ }
\accepted{ }
\cpright{AAS}{2000}
\shorttitle{The Cool ISM in S0 Galaxies}
\shortauthors{Welch \& Sage}

\begin{document}

\title{The Cool ISM in S0 Galaxies.  I. A Survey of Molecular Gas}

\author{Gary A. Welch}
\affil{Saint Mary's University}
\affil{Halifax, Nova Scotia}
\email{gwelch@orion.stmarys.ca}
\and
\author{Leslie J. Sage}
\affil{University of Maryland}
\affil{College Park, Maryland}
\email{lsage@astro.umd.edu}

\begin{abstract}

Lenticular galaxies remain remarkably mysterious as a class. Observations
to date have not led to any broad consensus about their origins, properties
and evolution, though they are often thought to have formed in one
big burst of star formation early in the history of the Universe, and
to have evolved relatively passively since then.
In that picture, current theory predicts that stellar evolution returns 
substantial quantities of gas to the interstellar medium; most is ejected 
from the galaxy, but significant amounts of cool gas might be retained.  
Past searches for that material, though, have provided unclear results. 
We present results from a survey of molecular gas in a volume-limited
sample of field S0 galaxies, selected from the Nearby Galaxies Catalog.
CO emission is detected from 78 percent of the
sample galaxies.  We find that the molecular gas is almost always located 
inside the central few kiloparsecs of a lenticular galaxy, meaning that in
general it is more centrally concentrated than in spirals. We combine our data 
with HI observations from the literature to determine the total masses of cool 
and cold gas.  Curiously, we find that, across a wide range of luminosity, the 
most gas rich galaxies have $\sim$10 percent of the total amount of gas ever 
returned by their stars.  That result is difficult to understand within the 
context of either monolithic or hierarchical models of evolution of the
interstellar medium.

\end{abstract}


\section{Introduction}

In broad terms, there are two very different ways in which galaxies could
have formed. The first, which we shall refer to as the monolithic formation 
idea, is that a large cloud of primordial gas collapsed rapidly to produce
either an elliptical or lenticular galaxy, or perhaps the spheroidal bulge of 
a spiral galaxy \citep{lar74,ari87,bow92}.
The main advantage of monolithic formation is that it seems consistent with the
observational facts that most such systems have relatively little angular
momentum and that they contain mostly old stars.  

The second way galaxies might have formed is through the hierarchical
accumulation of smaller clumps of gas and stars into larger structures that
became the galaxies we see today \citep{whi78,whi91,kau93,bau96}.
Within that view, there would be a continuum of mergers between clumps of
varying masses and angular momenta.  Some mergers would lead to bursts of
massive star formation, while others would not. Galaxy formation would be 
drawn out over a long period, perhaps continuing to the present epoch, where 
we still
see mergers and violent interactions between galaxies, along with the possible
formation of tidal dwarfs. The red colours of today's E and S0 galaxies would
be explained if most of the stars were produced in the original merger
participants.

Determining the observational properties of the different classes of galaxies
offers a way to discriminate between the possibilities. Until the IRAS
database
was available, the conventional wisdom was that elliptical and lenticular
galaxies had little or no gaseous interstellar medium -- this was reinforced
by the general paucity of HI emission.  For example, \citet{war86} and
\citet{brg92} find that most S0s have $M(HI)/L_B \lesssim 0.1$ in solar
units, but with large scatter, which is 3-6 times lower than spirals
\citep{rob75}. The surveys of atomic gas in early-type galaxies would of
course miss the molecular content, which we now know dominates the ISM near 
the
centers of spirals. The IRAS results motivated a few surveys of CO emission
(\citealp{sag89}; \citealp{thr89}; \citealp[hereafter WH]{wh89};
\citealp{lee91}; \citealp{wch95}). While the CO surveys were much less
complete in terms of sample size than those for HI, the general conclusion
was that in the IRAS-selected galaxies the interstellar medium -- at least 
near
the centers -- was not too different from that in normal, isolated spirals.
Comparison of the molecular and far-infrared properties showed that the 
process
of star formation in lenticulars is broadly similar to that in field spirals.

It seems unlikely, however, that the gaseous contents of the galaxies selected
using far-infrared criteria are representative of the bulk of lenticulars,
because most normal S0s have little or no far-infrared emission.
\citet{tan94} looked at 10 galaxies lacking optical emission lines, to
investigate a more typical sample. Their detection of six sources led them to
suggest that most normal S0s contain significant amounts of cool, dense gas,
which for unexplained reasons is not producing massive new stars.

Although S0s are dominated by $\sim 10^{10}$ year old stars, their low gas
content cannot merely be due to early, very efficient astration, where the gas
is simply locked into stars or blown out by winds from massive stars and
supernovae produced during a huge initial starburst. In the simplest
(closed box) view of galaxy evolution, one expects to find gas that has
subsequently been returned to the ISM by evolving low-mass stars, and the
latter contribution should be very large.  In general, the total mass of returned
gas M$_r$ scales with the present blue luminosity as M$_r$ = KL$_B$, with both
quantities in solar units.  \citet[hereafter FG]{fg76} adopt a reasonable return
rate for a 10 Gyr old population, that is, the product of an assumed planetary
nebula formation rate and mass loss by stars near 1 $M_{\sun}$.  Assuming the
rate to be constant over the galaxy lifetime implies that recycled gas alone
should presently give K$\sim 0.1$. More complex models of gas return within the
monolithic galaxy evolution scenario (e.g. \citealp{cio91,ken94,brm97}) predict
values more than an order of magnitude larger, owing to the early contributions
of massive stars during and after the spike of initial star formation.

In summary, there is very little understanding of how much gas is present in
field S0s, of its properties, and of how these properties might have arisen. To
improve this situation, and we hope thereby our understanding of what S0
galaxies are, has been the primary motivation for this study.

\section{The Sample}

The initial sample comprises all entries in the Nearby Galaxies Catalog
\citep{tul88} having type -3 to 0 inclusive, declination $\geq -10\arcdeg$, and
distance $<$ 20 Mpc.  We have omitted known peculiar galaxies, cases of 
probable
interaction (based on inspection of the field on POSS prints), members of the
Virgo cluster, and objects assigned class Sa in the Revised Shapley Ames
Catalog \citep[RSAC]{san86} or the Carnegie Atlas \citep{san94}.  The final
sample of 27 galaxies is listed in Table 1.  All but six of our sample are
contained in the RSAC.  Most (24/27) are classified as either S0, S0/Sa, or as
their barred counterparts.

Our final sample is free of the FIR bias inherent in earlier S0 surveys, but
still contains the biases of the Nearby Galaxies Catalog.  The most obvious of
these are that optically small galaxies
(diameter $<$ 1.5 $\arcmin$ -2 $\arcmin$ on the POSS) are excluded, and 
HI-poor
systems are under-represented.  Nonetheless, we believe that Table 1
constitutes the best sample now available for probing the cool ISM in 
normal S0
galaxies within low-density environments.  We report here on a single-antenna
survey of CO in these galaxies; a complementary HI study is underway.

In addition to its lack of FIR bias, the present work departs in two
important
ways from published ones, which have been essentially flux-limited, and have
usually looked only at the center of the galaxy \citep[see, however,][]{tac91}.  
First, we define a target
sensitivity for each galaxy using the Faber-Gallagher relation for an assumed
age of 10 Gyr to compute an expected ISM mass, $M_e$.  The goal is to 
provide a
5-sigma detection, or upper limit, of 0.02$M_e$ per pointing.  Since $M_e$
scales with the total blue luminosity, long integrations are needed for faint
galaxies.  Secondly, we observe additional points (along the major axis) in 
the
larger galaxies, even if no emission is detected at the center, as there are
several early-type galaxies with molecular gas outside their centers
\citep{tac91,sag94,ger94}.

\section{Observations and Data Reduction}

Spectra of the J=2-1 and J=1-0 transitions of CO in 20 galaxies were
obtained in October 1998 with the IRAM 30 meter telescope.  The NRAO 12 meter
telescope \footnote{The National Radio Astronomy Observatory is a facility of
the National Science Foundation operated under cooperative agreement by
Associated Universities, Inc.} was used in November 1998, January 2001, and in
February and April 2002 to observe the
remainder of the sample in the J=1-0 transition, and also the galaxies having
the largest angular sizes.  The IRAM telescope main beam is assumed to have
FWHM = 20.9$\arcsec$ and 10.4$\arcsec$ in the two transitions, whereas we take
FWHM = 55$\arcsec$ for the 12 meter telescope at CO (J=1-0).  Sufficiently
sensitive CO observations have been published for NGC 404, 7013, and 7077
\citep[WH;][]{sag89,sag90,sag92,tan94,yng95}, so we have not observed these 
galaxies.

The IRAM observations employed beam switching in azimuth with a throw of
240$\arcsec$ at 1 Hz.  Dual polarization measurements were
obtained simultaneously in both transitions, and directed both to
filterbanks and autocorrelator backends.  The filterbanks provided spectral
windows 1332 km s$^{-1}$ and 666 km s$^{-1}$ wide for
J=1-0 and J=2-1, respectively, with resolutions of 2.60 km s$^{-1}$ and
1.30 km s$^{-1}$ wide, respectively.  Corresponding values for the
autocorrelators were 1326 km s$^{-1}$ and 666 km s$^{-1}$, and 3.25 km 
s$^{-1}$
and 1.63 km s$^{-1}$, respectively.  Telescope pointing and focus were checked
about every 2 hours using a bright quasar or the planet Mars.  Over the
observing session, median corrections to the pointing model in azimuth and
altitude were 0.9$\arcsec$ and 1.4$\arcsec$, respectively.  System
temperatures were typically 300-400 K for J=1-0 and 600-700 K for J=2-1.

An initial 30m spectrum was obtained at the center of the target.  Additional
positions along the major axis were observed in 10 arcsecond steps, unless
either no emission was seen or it appeared that the J=2-1 transition was at
least as strong as J=1-0.  Because previous observers have generally reported
 \footnote{Following custom, I$_{21}$ and I$_{10}$ represent
the integrated intensities (area under the line) of the two CO transitions}
I$_{21}$/I$_{10}<$ 1 (see below), the latter case was assumed to result from
beam filling effects produced by a source smaller than the J=2-1 beam.

The large beam of the 12m telescope is useful in searching for extended gas
in larger galaxies.  Beam switching in azimuth was also
employed at that telescope, using a throw of 180$\arcsec$.  Filterbanks
collected data in both polarizations, providing windows 1330 km s$^{-1}$ wide
and individual channel widths of 5.2 km s$^{-1}$.  System temperatures were
usually 400-500 K.  Pointing and focus checks were made at roughly 2-hour
intervals, using Mars, Saturn, Jupiter, and several bright quasars.  Galaxies
were observed on center and at 55 arcsecond steps along the major axis.  The
observations typically encompass $\sim$ 0.5D$_{25}$ and $\sim$ 0.1D$_{25}$
along the major axis, for the 12m and 30m telescopes, respectively, making the
present survey unique for its extended spatial coverage.

The CLASS software package was used for data reduction.  Occasional bad
channels were first removed in the individual raw scans, which represented
4 and 6 minute integrations for the 30m and 12m telescopes, respectively.  All
scans from a given pointing were then summed and Hanning smoothed to typical
resolutions of 13.0 and 20.8 km s$^{-1}$ for the 30m and 12m telescopes,
respectively.  Other resolutions are noted in Tables 2 and 3.

Line windows were chosen as discussed below or in footnotes to Tables 2 and 3,
and linear baselines fitted to the remainder of the spectrum.  Second order
baselines were subtracted in a few cases, as noted in the tables.  Data were
obtained on the main beam temperature scale from the 30m telescope, and on the
T$_{R}^{*}$ scale at Kitt Peak.  The latter was transformed to T$_{mb}$
using a
corrected main beam efficiency of $\eta{_m}^{*}$ = 0.82.

When emission was not clearly detected, we defined spectral-line windows for
upper limit estimates using published kinematics of atomic gas;  otherwise,
the
optical rotation curve was employed.  In only two cases, NGC 4143 and NGC 
4346,
were we unable to find any guiding observations;  for these galaxies we used
windows 400 km s$^{-1}$ wide, centered on the optical velocity, as that
represents the linewidth of a 'typical' galaxy in our sample.

\section{Results}

\subsection{CO In S0 Galaxies}

The final summed spectrum at each center pointing is presented in Figure 1.  
Integrated intensities on the T$_{mb}$ scale from 30m and 12m
observations appear in Tables 2 and 3, respectively.  Standard deviations,
which serve as 1-sigma upper limits for non-detections, are calculated
following the prescription of \citet{sag90}.  The rms noise in the binned 
spectrum,
required for that procedure, is given in the table.  Observed positions are
identified by specifying the offsets in arcseconds from the coordinates given
in Table 1. 

Excluding objects already known to contain CO (NGC 404, 7013, 7077), we have
detected 75 percent (18/24) of our sample with a formal confidence of
3$\sigma$ or higher, 42 percent (10/24) with 5$\sigma$ or higher.  The 
detection rate for the entire sample is 78 percent.
Only the J=2-1 transition was detected in NGC 2787, 3115, 3384, and 3990;
although formally significant at 3.6$\sigma$ and 4.8$\sigma$, respectively,
the lines in NGC 3115 and NGC 3990 are complex and not convincing.
Emission (at 4.9$\sigma$) from NGC 4026 was detected only by the 12m telescope
at offset (2, -55).  In the following summary, we include these galaxies among
our detections; better data are clearly needed for them. 

Within our admittedly small sample of 27 objects, types S0 and S0/a are
detected with similar frequency (11/15 and 6/8, respectively).  Somewhat 
surprisingly, S0s with
optical dust features and those without are also detected at about the same
rate:  7/10 galaxies classified S0$_1$ or S0$_{1/2}$ (or as their barred
counterparts) are detected, and 4/6 of those classified as S0$_2$, S0$_{2/3}$
or S0$_3$.  Finally, SB0 galaxies are somewhat less likely to be detected (2/4) 
than S0 galaxies (8/10), though high inclinations might obscure bars in several 
of the latter.  In summary, except for a possible anti correlation with the 
presence of a bar, our sample does not show a relation between the gross 
optical properties of S0 galaxies and their detectability in CO emission.

The present observations can be used to estimate typical source sizes.  The
IRAM 30m data show that I$_{21}$/I$_{10}>$1 in 7 galaxies where only the
centre pointing was observed.  Even when the gas is quiescent, large intensity 
ratios can result when the source size is comparable to or smaller than the
J=2-1 beam.  Assuming that to be the
case, we derive a median linear size of 0.7 kpc for those 7 galaxies.  An
additional 7 galaxies have multiple-pointing 30m data showing
strong radial intensity declines which indicate that we have seen most of the
total CO emission. (This conclusion is less clear for NGC 4460; nonetheless,
we include this galaxy in the discussion).  We estimate the source size in
those cases by summing the maximum separation between pointings with 
detections and adding the FWHM of the appropriate telescope beam, finding a
median of 2.8 kpc, or 0.32D$_{25}$.  Therefore, our work reinforces the view
\citep[e.g.][]{tac91} that most of the molecular gas in S0 galaxies is near
their centres.

\subsection{Comparison of 2-1 and 1-0 emission}

Line ratios (i.e. ratios of integrated intensities in two transitions) are 
used to investigate the physical state of molecular gas, though the 
interpretation is complicated because of the different ways the telescope beams 
couple to the source, and by the inevitable mixing of emission from
dense cloud cores and more diffuse surrounding molecular gas.  The few published
multi-transition CO studies of early-type galaxies \citep[e.g.][]{wh90,sag93,
li93,wik97} report 0.5 $\lesssim$I$_{21}$/I$_{10}\lesssim$1,
which is assumed to indicate the gas is cool (T$_{ex} \sim$10K), optically
thick and generally quiescent.  

Without maps of our sources we cannot correct for different beam coupling.  
Instead, we compare in Figure 2 the 30m telescope results with 
predictions which assume the same radiation temperature in both
transitions, and that the source is either a point (top line) or that it
uniformly fills both beams.  The factor of 4 vertical difference between the 
lines is therefore the ratio of beam areas.  The results
shown in the figure are broadly consistent with previous studies.

\subsection{Molecular masses from the CO data}

Total H$_2$ masses for all detected galaxies, or their upper limits, are 
presented in Table 4.  Whenever possible we use the relation given by 
\citet{wel96} with a conversion constant of
2.3 $\times$ 10$^{20}$ cm$^{-2}$(K km s$^{-1}$)$^{-1}$.  
Uncertainties and upper limits are obtained by adding in quadrature
the appropriate uncertainties from individual pointings.  Molecular masses for
NGC 2787, 3115, 3384, 3990 are estimated from the J=2-1 line intensities,
assuming I$_{21}$/I$_{10}$=0.5 because, except perhaps for NGC 2787, no 
J=1-0 emission is seen.

We transform to our assumptions the published mass estimates for 6 galaxies
observed by others and present them in Table 5.  Values obtained by ourselves
from the 30m and 12m telescopes show pleasing consistency, which is generally
present throughout the results contained in Tables 2-4.  Agreement with
other studies is usually satisfactory, though the values reported by the NRAO 
12m and Nobeyama 45m telescopes for NGC 7457 are the inverse of what one 
might expect from comparing the two beam sizes.  Another 
noteworthy disparity is the factor of 2 
difference between 30m observations towards the centre of NGC 5866.  Most of the
disagreement comes from the integrated intensities, which we believe are both
on the T$_{mb}$ scale but which differ by a factor of 1.6.  Inspection of
Table 2 shows that I$_{10}$ changes by large amounts between positions 
separated by only $\sim$0.5$\times$FWHM or $\sim$10$\arcsec$ near the centre of 
the galaxy, leading us to suspect that telescope pointing is the culprit, as
previously suggested by \citet{sai91} for a different set of galaxies.

In Table 6 and Figure 3 we present estimates of the total cool gas content of
galaxies in our sample, obtained by combining molecular and atomic hydrogen 
masses from Table 4 and from the literature.  The table also shows the
predicted mass of gas returned by stars according to the analytical expression 
of \citet{cio91}, i.e. the case K = 1.2 as discussed below.  Together, 
Table 6 and Figure 3 encapsulate the major results of our study.  We now 
turn to the question of whether they support current ideas of the nature of 
lenticular galaxies.

\section{Origin and fate of the ISM in lenticular galaxies}

\subsection{Paradigms for S0 galaxy formation}

It has generally been thought that S0 galaxies formed most of their stars more
or less monolithically 10 Gyr or more ago, because of their typically red
colours and late-type stellar spectra (e.g. \citealp{bot90,sch92,mun92}).  That
view has been reinforced by the paucity of atomic hydrogen \citep{war86},
relatively low levels of far-infrared emission \citep{bal89}, and by the almost
complete absence of optical dust lanes \citep{ebn88,ver88} and HII regions
\citep{kim89,bus93,pog93} associated with star formation in spiral galaxies.  In
the monolithic paradigm, a large cloud of primordial gas collapsed fairly rapidly
to produce the galaxy, which has evolved passively since then.  More rapid
collapses produced elliptical galaxies, in which disks are insignificant; less
rapid collapses led to gas-rich spirals \citep{roc88,bru93}.  But the simple
monolithic picture has come under increasing pressure over the last decade,
because the predicted bright, massive protogalaxies that would be the precursors
of today's ellipticals have not been observed at any redshift or spectral band
in the necessary numbers \citep{mad96,kau98}, although the nature of the
unidentified galaxies seen through their sub-millimeter emission remains unclear
\citep{bar99,lil99}.  Those deficiencies have stimulated investigations of the
hierarchical formation idea, whereby galaxies are assembled from smaller pieces
through a series of mergers \citep{whi78,whi91,kau93,bau96}.

We initially hoped that our data would reveal evidence to prefer one path over
the other.  But as we show below, the most startling observational result
remains difficult to understand within either picture.

Our most significant result is contained in Figure 3, which displays the total
mass of atomic and molecular gas versus blue luminosity.  The wide range in gas
content among galaxies of similar luminosity is immediately evident.  The most
striking aspect, though, is the upper mass cutoff, which extends over the
entire luminosity range spanned by our data, a factor of $\sim100$ in
luminosity and presumably in total mass.  Within that range, no galaxy contains
more than about 10 percent of the gas predicted by recent estimates of stellar
mass return (see below), shown as a solid line in the figure.  Stated
differently, even the most gas-rich S0s have lost 90 percent or more of all the
gas ever returned by their stars.

We find ourselves unable to use the data to distinguish clearly between the two
evolutionary paths. As we outline below, with a judicious tuning of parameters
it is possible to explain aspects of our results within either evolutionary
scenario, but in both cases the tuning is contrived and the correspondence with
the data incomplete.  Perhaps the difficulty arises from an assumption
underlying our work - that all S0 galaxies have evolved along very similar paths.
Van den Bergh (1990,1994) suggests that lenticulars comprise at least two
distinct
kinds of objects, one of these being the true S0s, i.e. galaxies forming the
transition between ellipticals and spirals.  Although it is not yet possible to
identify the pedigree of an individual galaxy, van den Bergh suggests that true S0s
are hiding among brighter, rounder galaxies, and that they tend to have
\citet{hog93} sub-type I, I/S, or S.  We have tried to test that suggestion by
searching for some trend in optical properties among the galaxies in our
sample.  We have asked, for example, if the upper cutoff might be delineated by
true S0s, but have not found any evidence to support the idea.  While a wide
range of gas content is certainly consistent with the view of S0s as a
heterogeneous collection, the upper cutoff probably is not.

In the following sections we attempt to lay out the issues for the two galaxy
formation paradigms, without providing a coherent explanation for our results.

\subsection{Monolithic model predictions}

Faber and Gallagher realized that any isolated, passively evolving galaxy which
converted almost all of its gas into stars early in its existence would over a
Hubble time build up an embarrassingly massive ISM, unless most of the returned
gas could be disposed of.  The simple picture of stellar mass return used by FG
has been extensively revised, most recently by \citet{cio91}, and
\citet{brm97}, without qualitatively altering the original conclusion.  We remind
the reader of the model predictions that the mass of returned gas is given by
M$_r$ = KL$_B$, with all the relevant and variable physics being subsumed into
the factor K.

The study by FG includes only gas return from stars with mass
$\sim$1M$_{\sun}$.  When updated with more recent planetary nebula formation
rates \citep{pei93}, it predicts that K = 0.07, shown by a dashed line in
Figure 3.  The assumption underlying the FG approach is that gas returned to the
ISM during the first few Gyr of the  galaxy's existence was blown into
intergalactic space by type II supernovae.

More complex descriptions of stellar gas return, which include contributions
from a range of masses described by Salpeter or Miller-Scalo initial mass
functions, predict larger values of K, i.e. K$\sim$1.  Those descriptions have
been incorporated in models of ISM evolution which attempt to explain the X-ray
emission from plasma within the deep potential wells of optically luminous E and
S0 galaxies.  The assumed masses are therefore considerably larger than
appropriate for most of the galaxies in our sample.  With that caveat in mind,
the studies most relevant for interpreting our data are Brighenti \& Mathews's
(1997) work on flattened ellipticals, and D'Ercole \& Ciotti's (1998)
investigation of lenticular galaxies.

\citet{brm97} and \citet{dec98} take the initial state of the ISM to
be a hot, outflowing wind - the result
of heating by type II supernovae in the initial starburst. If all gas returned
after the first $\onehalf$ Gyr is retained, the analytic approximation of
\citet{cio91} yields K = 1.2 for galaxies assumed to be 10 Gyr old today,
implying the presence of a massive ISM.  The amount of gas actually retained,
however, is very sensitive to the depth and shape of the gravitational
potential, and also to the assumed rate of energy input from type Ia
supernovae. In all cases, though, the results are qualitatively consistent
with our observations: most of the gas ever returned to the ISM is lost, much
of it in the first 1 Gyr of the galaxy's life.

Models of more massive galaxies show the development of central cooling flows,
also termed decoupled inflows, which accumulate 10$^{9-10}$M$_{\sun}$ of cool gas.
No attempt has yet been made to explore the fate of the cooling gas.  The amount
of cool gas expected at the present time depends on when the initial outflow
reverses, i.e. on the assumed SN Ia rate for a given gravitational potential.
Uncertainty in the supernova rate is a major impediment to understanding what
happens to the gas, but using rates consistent with recent work
\citep[e.g.][]{tur94,ari97,mat97} the models of D'Ercole \& Ciotti predict that
the most massive galaxies in the present sample, are likely to have developed
inflows after several Gyr.

Models are not yet available of flattened galaxies with M$_{tot}\sim3\times
10^{10}$M$_{\sun}$, which is more typical of our sample, but the trends make it
appear that, due to their low masses, few of our galaxies can have developed
inflows, so they should be essentially free of cool gas.  We therefore
anticipate that models with galaxy masses more appropriate to our sample will
predict that effectively no gas will be retained in the smaller
galaxies.  What is particularly striking, in the light of that expectation, is
that we see molecular gas even in some of the lowest-mass galaxies in our sample.
Figure 3 reveals that those galaxies have gas deficiencies equal to (and in some
cases much less than) the deficiencies of the most massive galaxies we have
observed - objects roughly 100 times more massive. Varying the energy equilibrium
of the gas(e.g. the SN Ia rate) can explain the range of gas mass at any
particular blue luminosity.  But it seems excessively contrived to explain the
upper cutoff by such parameter adjustments.

\subsection{Predictions of Hierarchical models}

The optical colours and spectra of S0 disks are consistent with them being a few
Gyr younger than the accompanying bulges \citep{cal83,bot90,fis96}, although most
observations refer to S0s in clusters.  Relatively young disks can be readily
explained by one of the central pillars of the hierarchical paradigm: spheroids
formed in major mergers can later acquire disk components from surrounding gas
reservoirs (Kauffmann et al. 1993; Steinmetz \& Navarro 2002).
While the evolution of the morphology and gas-content of such disks has not been
clarified by current hierarchical studies, the idea of gas infall over extended
periods seems to offer no ready explanation for the upper cutoff in Figure 3.

The fundamental reason why hierarchical models cannot yet address our results
is that the fate of the gas in a \emph{single} merger cannot be predicted
with confidence.  Interactions between galaxies bring in new gas.  They also
dispose of it in star formation, by blasting it into intergalactic space through
type II supernovae, or by tidal stripping.  The outcome therefore depends on
many parameters, including the structure of the colliding galaxies, the
encounter geometry, the prescription for star formation, and the effectiveness
of stellar feedback\citep[e.g.][]{bar91,mih92,how93,bek94,her95,mih96,bar96,
bk98a,bk98b,bk98c,rou01,bar02}.  It is again conceivable that judicious
parameter tuning could explain the observed range of gas deficiency. We doubt
the same is true for the upper cutoff.

\section{Summary}

We report a survey of molecular gas in a sample of 27 normal, field S0
galaxies
within 20 Mpc.  The present survey differs from earlier ones in that it is
volume-limited rather than IRAS flux-limited, looks for gas outside the
centres
of the galaxies, and has sensitivity limits based on fundamental
considerations
of stellar evolution - that aging stars return gas to the interstellar
medium.
Our results provide a unique basis for understanding the properties, origin,
and fate of dense gas in normal S0 galaxies residing outside of rich clusters.

We confirm earlier studies of different samples, finding that molecular gas
is common in S0 galaxies; 78 percent of our sample is detected in
either the J=1-0 or J=2-1 transition of CO.  The gas is centrally located in
regions with sizes up to 1-3 kpc;  we do not find significant amounts of gas at 
larger distances.

We have combined observations with published HI data to estimate the total
mass of cold gas in each galaxy.  The results of that exercise are 
presented in Figure 3, which contains two important results:  First, a wide
range of gas content is present at a fixed optical luminosity.  Second, an 
upper cutoff in gas content is present across the entire luminosity range 
spanned by our sample: -17 $\gtrsim$M$_B\gtrsim$-21.  Those findings, but 
especially the second, present a significant challenge for our ideas of how 
S0 galaxies have evolved. 

If S0s formed during a single monolithic collapse early in the history of
the Universe, and have subsequently evolved as \emph{closed systems}, then all 
now contain much less gas than expected from stellar mass loss; Figure 3 
indicates that 10$^{-4}\lesssim M_{obs}/M_{ret}\lesssim$10$^{-1}$.  That fact
alone, though, does not necessarily present a serious problem for models of ISM
evolution based on the monolithic formation paradigm.  The expected amount of
cool gas is extremely sensitive to poorly known
quantities such as the assumed rate of type Ia supernovae, and the shape and
depth of the gravitational potential well, so that a wide range of gas content
can be explained.  It is very hard to understand, however, how those poorly
known factors, and perhaps others, could conspire to produce the upper cutoff
of gas content so clearly evident in Figure 3..

Within the paradigm of hierarchical galaxy formation, S0 bulges may have formed
through major mergers, followed by the slow accretion of surrounding gas to form
their disks. Although that idea can explain optical observations which imply
relatively young ages for S0 disks, it might be difficult to understand how the
upper cutoff would arise from an extended period of gas accretion.

In summary, our study poses broad and significant questions for the two 
popular galaxy formation paradigms, and it is not yet clear which, if either, 
will be more successful at answering them.  It will certainly be helpful to 
substantially deepen our understanding of the properties of the cold gas, and 
we are embarking on projects for this purpose.

\acknowledgments

We are grateful to Dave Hogg and Luca Ciotti for their perceptive comments on 
early versions of this paper, and to Bill Mathews for insights into ISM 
evolution in galaxies.  Comments from the referee helped us to improve our
presentation.  This work was supported by a Canadian Natural Sciences and
Engineering Research Council research grant to GAW.

\pagebreak

\figcaption{Completely reduced spectra obtained by pointing the IRAM 30m (first 
5 pages) and NRAO 12m (following 3 pages) telescopes at the positions listed in 
Table 1 - offsets (0,0), except as indicated for NGC 4026. Heliocentric 
velocity and main beam temperature in K are plotted.  The 30m data are 
arranged with J=(1-0) spectra in the left column and J=(2-1) in the right 
column, as indicated at the top of each page.  All 12m spectra show the J=(1-0) 
transition.  Details of the reduction path are given in Section 3.}

\figcaption{CO intensities from the IRAM 30m Telescope at offset (0,0) (Table 2) 
are compared with predictions from assuming the same radiation temperature in 
both transitions, and that the sources are points (top line) or that they 
uniformly fill both beams (bottom line).  The factor of 4 vertical offset 
between the lines is essentially the same as the ratio of beam areas.  
One-sigma error bars are indicated for the most uncertain data.}

\figcaption{Cool gas masses versus blue luminosity for the galaxies in our 
sample.  Helium is included and solar units are used for both quantities.  
Points locate entries from Column 5 of Table 6.  They represent masses derived 
from detections in both CO and HI (bare points), from detections of one phase 
added to 3-$\sigma$ limits on the other phase (points with dashed lines), from 
the sum of 3-$\sigma$ limits (down arrows), or from CO detections when no HI 
observation exists (up arrows).  The lower ends of the dashed lines locate 
masses obtained from the measured phase (almost always CO).  The solid line 
gives the mass of returned gas predicted by the analytic approximation of 
\citep{cio91}, and includes everything returned after the first 500 Myr.  The 
dashed line shows the mass returned only by solar mass red giants over the past 
10 Gyr \citep{fg76}   }
\pagebreak
\begin{deluxetable}{lcccccc}
\tablecolumns{7}
\tablewidth{0pc}
\tablecaption{Properties of Galaxies in the Volume-limited Sample}
\tablehead{
\colhead{Name} & \colhead{RA} & \colhead{DEC} & \colhead{v$_{hel}$} &
\colhead{Type} & \colhead{D}  & \colhead{B${_T}{^o}$} \\
\colhead{} & \multicolumn{2}{c}{1950} & \colhead{km s$^{-1}$} &
\colhead{} & \colhead{Mpc} & \colhead{} }
\startdata
NGC 404  & 01:06:39.3 &  35:27:06 &  -35 &     S03(0) &  2.4 & 10.92 \\
NGC 936  & 02:25:04.7 & -01:22:42 & 1340 & SB02/3/SBa & 10.6 & 10.98 \\
NGC 1023 & 02:37:15.9 &  38:50:55 &  637 &    SB01(5) & 10.5 & 10.08 \\
NGC 2787 & 09:14:49.7 &  69:24:50 &  696 &      SB0/a & 13   & 11.61 \\
NGC 3115 & 10:02:44.7 & -07:28:31 &  670 &     S01(7)  &  6.7 &  9.74 \\\\

NGC 3384 & 10:45:38.7 &  12:53:41 &  735 &    SB01(5)  &  8.1 & 10.75 \\
NGC 3412 & 10:48:14.6 &  13:40:41 &  865 &  SB01/2(5)  &  8.1 & 11.34 \\
NGC 3489 & 10:57:40.0 &  14:10:15 &  708 &     S03/Sa  &  6.4 & 11.15 \\
NGC 3607 & 11:14:16.1 &  18:19:35 &  935 &     S03(3)  & 19.9 & 10.79 \\
NGC 3870 & 11:43:16.9 &  50:28:43 &  750 &       [S0]  & 17   & 13.47 \\\\

NGC 3941 & 11:50:19.4 &  37:15:55 &  928 &   SB01/2/a  & 18.9 & 11.25 \\
NGC 3990 & 11:55:01.0 &  55:44:15 &  696 &     S06/Sa  & 17   & 13.46 \\
NGC 4026 & 11:56:51.1 &  51:14:25 &  930 &   S01/2(9)  & 17   & 11.59 \\
NGC 4111 & 12:04:30.6 &  43:20:43 &  807 &     S01(9)  & 17   & 11.6  \\
NGC 4143 & 12:07:05.0 &  42:48:52 &  784 &  S01(5)/Sa  & 17   & 11.9  \\\\\\

NGC 4150 & 12:08:01.3 &  30:40:47 &  244 &  S03(4)/Sa  &  9.7 & 12.4  \\
NGC 4203 & 12:12:34.1 &  33:28:33 & 1086 &     S02(1)  &  9.7 & 11.61 \\
NGC 4251 & 12:15:36.8 &  28:27:11 & 1014 &     S01(8)  &  9.7 & 11.43 \\
NGC 4310 & 12:19:56.1 &  29:29:10 &  913 &    [SAB0?]  &  9.7 & 13.08 \\
NGC 4346 & 12:21:01.2 &  47:16:16 &  783 &    SB01(8)  & 17   & 12.17 \\\\

NGC 4460 & 12:26:20.2 &  45:08:21 &  528 &      S0/Sc  &  8.1 & 12.25 \\
NGC 4880 & 12:57:41.0 &  12:45:10 & 1491 & [E4/S01(4)  & 15.7 & 12.35 \\
NGC 5866 & 15:05:07.8 &  55:57:16 &  745 &     S03(8)  & 15.3 & 10.83 \\
NGC 7013 & 21:01:26.1 &  29:41:51 &  779 &     [S0/a]  & 14.2 & 10.7  \\
NGC 7077 & 21:27:27.5 &  02:11:39 & 1146 & [S0 pec ?]  & 13.3 & 13.94 \\\\

NGC 7332 & 22:35:01.2 &  23:32:16 & 1300 &   S02/3(8)  & 18.2 & 11.93 \\
NGC 7457 & 22:58:36.2 &  29:52:31 &  824 &     S01(5)  & 12.3 & 11.76 \\

\enddata

\tablecomments{Columns contain galaxy name, coordinates at epoch 1950,
heliocentric radial velocity, morphological type from the Carnegie Atlas or
Third Reference Catalog of Bright Galaxies (square brackets), distance in Mpc
from the Tully Catalog, and total corrected blue apparent magnitude from the
Third Reference Catalog of Bright Galaxies.}

\end{deluxetable}

\pagebreak
\begin{deluxetable}{lcccccccccccccc}
\tablewidth{0pc}
\tabletypesize{\footnotesize}
\rotate{}
\tablecolumns{15}
\tablecaption{Integrated Intensities from the IRAM 30m Telescope}
\tablehead{
\colhead{Name} & \colhead{} & \colhead{} & \multicolumn{2}{c}{1-0} &
\colhead{} & \multicolumn{2}{c}{2-1} & \colhead{} & \colhead{} & 
\multicolumn{2}{c}{1-0} & \colhead{} & \multicolumn{2}{c}{2-1}\\
\cline{4-5}  \cline{7-8}  \cline{11-12}  \cline{14-15}\\
\colhead{} & \colhead{offset} & \colhead{window} & \colhead{I$_{CO} \pm \sigma_I$} &
\colhead{rms} & \colhead{} & \colhead{I$_{CO} \pm \sigma_I$} & \colhead{rms} &
\colhead{} & \colhead{window} & \colhead{I$_{CO} \pm \sigma_I$} &
\colhead{rms} & \colhead{} & \colhead{I$_{CO} \pm \sigma_I$} & \colhead{rms} \\
\colhead{} & \colhead{$\arcsec,\arcsec$} & \colhead{km s$^{-1}$} &
\colhead{K km s$^{-1}$} & \colhead{mK}  & \colhead{} & 
\colhead{K km s$^{-1}$} & \colhead{mK}  & \colhead{} &
\colhead{km s$^{-1}$} & \colhead{K km s$^{-1}$} & \colhead{mK} & \colhead{} & 
\colhead{K km s$^{-1}$} & \colhead{mK}  }
\startdata

NGC 2787 & 0, 0     & 493-906 & 1.02 0.35 & 3.92     &   & 1.93 0.46 & 3.78  &  &
                     \nodata & \nodata & \nodata     &   & \nodata & \nodata      \\

NGC 3115 & -14, -17 & 360-450\tablenotemark{a} & 0.216 0.190 & 5.36 &  & 0.953 0.251
 & 6.81  &  &
                      360-960 & -0.423 0.725 &  6.02 &   & \nodata & \nodata      \\
                      
         &   0, 0   & 500-840 & \nodata  & \nodata   &   & 1.95 0.54 & 5.58  & &
                      360-960 & 0.232 0.584  &  4.85 &   & \nodata & \nodata       \\

         &  14, 17  & 870-960 & 0.023 0.191 & 5.37   &   & 0.320 0.233 &  6.32 &  &
                      360-960 & -1.42 0.67  & 5.55   &   & \nodata & \nodata      \\

NGC 3384 & 0, 0     & 535-935 & -0.796 0.260 & 3.00  &   & 2.49 0.56   & 4.75  &  &
                    \nodata & \nodata & \nodata      &   & \nodata & \nodata      \\

NGC 3412 & 0, 0     & 727-986 & 0.788 0.204 & 3.18   &   & 0.990 0.307 & 4.20  &  &
		    \nodata & \nodata & \nodata      &   & \nodata & \nodata      \\

NGC 3489 & 0, 0     & 521-879 & 3.49 0.39  &  4.21   &   & 4.86 0.41 & 3.97    &  &
		    \nodata & \nodata & \nodata      &   & \nodata & \nodata      \\		    

NGC 3607 & -9.5, 5.5 & 631-1238 & 2.59 0.37 & 3.16   &   & \nodata & \nodata   &  &
		       839-1221 & \nodata & \nodata  &   & 2.44 0.46  & 4.17      \\
		       
         & 0,   0    & 670-1227 & 8.74 0.97 & 4.16   &   & 9.85 0.56  & 2.35   &  &
         	     \nodata & \nodata & \nodata     &   & \nodata & \nodata      \\
         	     
         & 9.5, -5.5 & 634-1225 & 10.46 0.36 & 3.05  &   & \nodata &  \nodata  &  &
         	       645-1219 & \nodata  & \nodata &   & 21.62 0.49 & 1.84      \\
         	       
         & 19.1, -11.0 & 632-1161 & 5.35 0.45 & 4.14 &   & \nodata &\nodata    &  &
        		 670-845 & \nodata & \nodata &   & 2.55 0.31 & 5.08       \\

NGC 3870 & -4.7, -10 & 714-900 & 1.13 0.25 & 3.61\tablenotemark{b} &  & \nodata &
 \nodata   &  &
           	       736-880 & \nodata & \nodata   &   & 1.55 0.36  &
 8.13\tablenotemark{c}  \\
           	       
         & 0, 0      & 700-850 & 1.60 0.15 & 2.38\tablenotemark{b} &   & \nodata &
 \nodata  &  & 
          	       700-800 & \nodata & \nodata   &   & 2.46 0.24   &
 6.90\tablenotemark{c}  \\
          	       
         & 4.7, 10   & 701-817 & 0.894 0.120 & 2.33\tablenotemark{b} & & \nodata &
 \nodata  &  &
          	       708-797 & \nodata & \nodata   &   & 0.780 0.240 &
 7.17\tablenotemark{c}  \\

NGC 3941 & 0, 0  & 728-1128 & 0.729 0.295 & 2.41\tablenotemark{d} &  & 0.267 0.716 &
 4.30\tablenotemark{d}  &  &
		 \nodata & \nodata & \nodata     &   & \nodata & \nodata      \\

NGC 3990 & 0, 0  & 625-865  & 0.328 0.177 & 1.95\tablenotemark{d} & & 0.863 0.181 &
 1.81\tablenotemark{d}   &  &
		 \nodata & \nodata & \nodata     &   & \nodata & \nodata      \\

NGC 4026 & 0, 0  & 780-1080 & 0.183 0.206 & 2.02\tablenotemark{d} & & 0.072 0.588 &
 4.85\tablenotemark{d}   &  &
		 \nodata & \nodata & \nodata     &   & \nodata & \nodata      \\

NGC 4111 & 0, 0  & 607-1007 & 1.29 0.29 & 2.38\tablenotemark{d} & & 3.80 0.35 &
 2.09\tablenotemark{d}  &  &
		 \nodata & \nodata & \nodata     &   & \nodata & \nodata      \\

NGC 4143 & 0, 0  & 584-984 & 0.003 0.359 & 3.26\tablenotemark{b} & &  \nodata &
 \nodata  &  &
		   700-984 & \nodata & \nodata   &   & 1.17 0.91 & 6.07\tablenotemark{b}  \\

NGC 4150 & -7.5, 8 & 25-320 & 3.04 0.26 & 3.54   &   & 1.28 0.70 & 8.03  &  &
		   \nodata & \nodata & \nodata   &   & \nodata & \nodata      \\
		   
         & 0, 0  & 22-362 & 6.61 0.26 & 3.41     &   & 4.65 0.36 & 3.66  &  &
         	 \nodata & \nodata & \nodata     &   & \nodata & \nodata      \\
         	 
         & 7.7 -8 & 22-362 & 6.74 0.39 & 4.97    &   & 8.96 0.48 & 4.90   &  &
         	  \nodata & \nodata & \nodata    &   & \nodata & \nodata      \\
         	  
         & 15, -16 & 180-364 & 2.50 0.31 & 5.66  &   & \nodata & \nodata  &  &
          	     262-336 & \nodata & \nodata &   & 0.798 0.280 & 8.57     \\

NGC 4203 & 0, 0    & 868-1300 & 1.88 0.21 & 1.59\tablenotemark{d} & & \nodata &
 \nodata  &  &
          	     891-1261 & \nodata & \nodata &   & 4.52 0.35 & 3.24      \\

NGC 4251 & 0, 0    & 914-1114 & 0.094 0.195 & 3.51 &   & -0.346 0.417 & 6.79  &  &
		   \nodata & \nodata & \nodata     &   & \nodata & \nodata      \\

NGC 4310 & -26, 20 & 881-1024 & 0.408 0.270 & 4.34 &   & 0.744 0.290 & 5.94   &  &
		   \nodata & \nodata & \nodata     &   & \nodata & \nodata      \\
		   
         & -17, 14 & 881-1024 & 3.12 0.22 & 4.78   &   & 2.48 0.37 & 7.63    &  &
         	   \nodata & \nodata & \nodata     &   & \nodata & \nodata      \\
         	   
         &  -9, 7  & 853-1005 & 5.70 0.23 & 4.88   &   & 6.86 0.39 & 7.62    &  &
         	   \nodata & \nodata & \nodata     &   & \nodata & \nodata      \\
         	   
         & 0, 0    & 813-996  & 6.69 0.31 & 5.86   &   & 7.17 0.37 & 6.35    &  &
         	   \nodata & \nodata & \nodata     &   & \nodata & \nodata      \\
         	   
         &  9, -7  & 799-993  & 3.60 0.24 & 4.45   &   & 2.91 0.36 & 6.05    &  &
         	   \nodata & \nodata & \nodata     &   & \nodata & \nodata      \\
         	   
         & 17, -14 & 813-929  & 0.847 0.184 & 4.53 &   & 0.396 0.290 & 6.71  &  &
         	   \nodata & \nodata & \nodata     &   & \nodata & \nodata      \\

NGC 4346 & 0, 0 & 583-983 & 0.407 0.413 & 3.77\tablenotemark{b} & & -0.196 0.726 &
 6.90\tablenotemark{c}  &  &
		  \nodata & \nodata & \nodata     &   & \nodata & \nodata      \\

NGC 4460 & -11, -19 & 390-500 & 1.10 0.25 & 5.04\tablenotemark{b} & & 0.234 0.274 &
 5.22\tablenotemark{b}  &  &
		    \nodata & \nodata & \nodata     &   & \nodata & \nodata      \\
		    
         & -6, -10  & 410-544 & 0.988 0.218 & 3.85\tablenotemark{b} & & 0.622 0.354 &
 5.97\tablenotemark{b}  &  &
         	    \nodata & \nodata & \nodata     &   & \nodata & \nodata      \\
         	    
         & 0, 0     & 402-564 & 2.12 0.20 & 3.16\tablenotemark{b} & & 0.676 0.315 &
 4.63\tablenotemark{b}  &  &
         	    \nodata & \nodata & \nodata     &   & \nodata & \nodata      \\
         	    
         &  6, 10  & 454-584 & 2.24 0.25 & 4.39\tablenotemark{b} & & -0.339 0.305 &
 5.23\tablenotemark{b}  &  &
         	   \nodata & \nodata & \nodata     &   & \nodata & \nodata      \\
         	   
         & 11, 19  & 514-565 & 0.560 0.118 & 3.54\tablenotemark{b} & & 0.588 0.193 &
 5.67\tablenotemark{b}  &  &
         	   \nodata & \nodata & \nodata     &   & \nodata & \nodata      \\

NGC 4880 & 0, 0    & 1342-1413 & 0.823 0.110 & 3.52 &   & 0.999 0.110 & 4.72  &  &
		   \nodata & \nodata & \nodata      &   & \nodata & \nodata      \\

NGC 5866 & -35, 27 & 479-606 & 0.441 0.170 & 4.02 &   & 0.810 0.360 & 7.94  &  &
		   \nodata & \nodata & \nodata    &   & \nodata & \nodata      \\
		   
         & -26, 20 & 479-606 & 1.68 0.24  & 5.64  &   & 1.04 0.28  & 6.27  &  &
         	   \nodata & \nodata & \nodata    &   & \nodata & \nodata      \\
         	   
         & -17, 14 & 466-641 & 4.57 0.19  & 4.01  &   & 4.03 0.33  & 6.18  &  &
         	   \nodata & \nodata & \nodata    &   & \nodata & \nodata      \\
         	   
         & -9, 7   & 458-752 & 10.97 0.28  & 3.95 &   & 12.53 0.65  & 7.59  &  &
         	   \nodata & \nodata & \nodata    &   & \nodata & \nodata      \\
         	   
         & 0, 0   & 432-1051 & 21.61 0.52 & 4.23  &   & \nodata & \nodata  &  &
          	    455-1014 & \nodata & \nodata  &   & 18.79 0.69 & 2.89      \\
          	    
         &  9, -7 & 410-1087 & 28.06 0.37 & 2.98  &   & \nodata & \nodata  &  &
          	    472-1040 & \nodata & \nodata  &   & 32.79 0.83 & 3.23      \\
          	    
         & 17, -14 & 708-1047 & 13.83 0.35 & 4.31 &   & \nodata & \nodata  &  &
          	     797-1022 & \nodata & \nodata &   & 13.69 0.50 & 7.43      \\
          	     
         & 26, -20 & 816-1048 & 5.01 0.29 & 4.46  &   & \nodata & \nodata  &  &
          	     834-1027 & \nodata & \nodata &   & 6.94 0.36 & 5.99       \\
          	     
         & 35, -27 & 866-1100 & 1.68 0.35 & 5.38  &   & \nodata & \nodata  &  &
          	     854-1014 & \nodata & \nodata &   & 0.449 0.390 & 7.44     \\
          	     
         & -9, -7\tablenotemark{e} & 465-1042 & 6.27 0.82 & 6.63 &  & \nodata &
 \nodata  &  &
        			     521-1036 & \nodata & \nodata & & 2.94 0.93 & 5.00       \\
        			     
         &  9, 7  & 455-1061 & 15.55 0.82 & 6.59  &   & \nodata & \nodata  &  &
           	    479-1037 & \nodata & \nodata  &   & 25.15 1.21 & 5.03       \\

\enddata

\tablecomments{Columns contain the galaxy name, offset in arcseconds (RA,DEC)
from coordinates in Table 1, location of line window, integrated intensity in
the line window and its formal standard deviation, and rms channel noise in 
the smoothed spectrum used for the measurements.}

\tablenotetext{a}{First window at each offset brackets optical rotation
curve at that position (Refer to notes for Table 3).  Second window spans
entire rotation curve.}

\tablenotetext{b}{20.8 km s$^{-1}$ channel width.}

\tablenotetext{c}{10.4 km s$^{-1}$ channel width.}

\tablenotetext{d}{26.0 km s$^{-1}$ channel width.}

\tablenotetext{e}{This and following pointing is along the minor axis.}
\end{deluxetable}
\begin{deluxetable}{lccccc}
\tablecolumns{6}
\tablewidth{0pc}
\tablecaption{Integrated Intensities from the NRAO 12m Telescope}
\tablehead{
\colhead{Name} & \colhead{} & \colhead{} & \multicolumn{3}{c}{1-0} \\
\cline{4-6} \\
\colhead{} & \colhead{offset} & \colhead{window} & \colhead{I$_{CO}$} &
\colhead{$\sigma_I$} & \colhead{rms} \\
\colhead{} & \colhead{$\arcsec,\arcsec$} & \colhead{km s$^{-1}$} &
\colhead{K km s$^{-1}$} & \colhead{K km s$^{-1}$} & \colhead{mK}  }
\startdata

NGC 936  & -39, 39 & 1490-1580\tablenotemark{a} & -0.021 & 0.134 & 2.98  \\
          &   0, 0  & 1190-1490 & 0.058  & 0.408 & 4.52  \\
          &   39, -39 & 1100-1190 & -0.037 & 0.181 & 4.02 \\

NGC 1023 & -220, -12 \tablenotemark{b} & 430-760  & 0.087  & 0.486 & 5.07  \\
          & -220, -12 & 400-1000 & 0.792  & 0.721 & 4.74  \\
          & -165, -9  & 430-760  & 0.217  & 0.224 & 2.33  \\
          & -165, -9  & 400-1000 & \nodata\tablenotemark{c} & 0.357 & 2.33  \\
          & -110, -6  & 430-800  & -0.366 & 0.395 & 3.80  \\
          & -110, -6  & 400-1000 & -0.195 & 0.519 & 3.38  \\
          &  -55, -3  & 530-800  &  0.283 & 0.292 & 3.46  \\
          &  -55, -3  & 400-1000 & -0.141 & 0.476 & 3.11  \\
          &    0, 0   & 530-800  &  -0.014 & 0.149 & 1.77  \\
          &    0, 0   & 400-1000 &  0.449 & 0.235 & 1.54  \\
         &   55, 3   & 530-800  &  0.056 & 0.251 & 2.97  \\
         &   55, 3   & 400-1000 &  1.24  & 0.45  & 2.91  \\
         &  110, 6   & 530-850  &  1.02  & 0.53  & 5.62  \\
         &  110, 6   & 400-1000 &  1.10  & 0.80  & 5.19  \\
         &  165, 9   & 600-850  &  0.080 & 0.221 & 2.75  \\
         &  165, 9   & 400-1000 & -0.911 & 0.381 & 2.49  \\
         &  220, 12  & 600-850  & -0.338  & 0.152  & 1.90  \\
         &  220, 12  & 400-1000 & -0.293  & 0.275  & 1.79  \\

NGC 2787 &  0, 0    &  580-690  & 0.355  &  0.072  & 1.37  \\

NGC 3115 &  -71, -84 & 360-450\tablenotemark{a}  & -0.312 & 0.238 & 5.31  \\
         &  -35, -42 & 360-450  &  0.029 & 0.129 & 2.87  \\
         &    0, 0   & 360-960  & -0.335 & 0.582 & 3.80  \\
         &   35, 42  & 870-960  & -0.021 & 0.162 & 3.60  \\
         &   71, 84  & 870-960  & -0.111 & 0.188 & 4.19  \\

NGC 3384 &    0, 0   & 535-935  & -0.136 & 0.406 & 3.66  \\
         &   44, 33  & 535-935  & -0.547 & 0.312 & 2.94  \\

NGC 3412 &    0, 0   & 727-986  & -0.420 & 0.250 & 3.03  \\

NGC 3607 &    0, 0   & 634-1221 &  8.72  & 0.96  & 6.42  \\

NGC 3870 &    0, 0   & 635-937  &  1.50  & 0.36  & 4.11  \\

NGC 3941 &    0, 0   & 728-1128 &  0.620 & 0.197 & 1.79  \\

NGC 4026 &   -2, 55  & 630-920\tablenotemark{a}  &  0.532 & 0.241 & 2.73  \\
         &    0, 0   & 630-1180 &  0.043 & 0.325 & 2.29  \\
         &    2, -55 & 900-1180 &  1.03  & 0.21  & 2.47  \\

NGC 4111 &    0, 0   & 607-1007 & -0.793 & 0.510 & 4.64  \\

NGC 4251 &    0, 0   & 864-1164 & -0.336 & 0.227 & 2.52  \\

NGC 4310 &  -43, 34  & 881-1024 &  0.462 & 0.310 & 5.28  \\
         &    0, 0   & 813-996  &  1.56  & 0.71  & 1.07  \\
         &   43, -34 & 813-929  & -0.088 & 0.230 & 4.48  \\

NGC 4346 &    0, 0   & 583-983  & -0.475\tablenotemark{d} & 0.324 & 2.94  \\

NGC 4460 &    0, 0   & 392-586  &  0.676 & 0.200 & 3.22  \\
         &   35, 42  & 434-606  &  0.948 & 0.340 & 5.16  \\

NGC 5866 &  -86, 68  & 479-606  & -0.190 & 0.110 & 4.25  \\
         &  -43, 34  & 479-606  & -0.250 & 0.130 & 4.85  \\
         &    0, 0   & 455-1014 &  5.14  & 0.49  & 3.14  \\
         &   43, -34 & 931-1062 &  0.767 & 0.100 & 3.68  \\
         &   86, -68 & 931-1062 & -0.462 & 0.110 & 3.95  \\

NGC 7332 &  -23, 50  & 1100-1500& -1.17  & 0.31  & 2.81  \\
         &    0, 0   & 1100-1500& 0.186  & 0.089 & 0.81  \\
         &    0, 0   & 1370-1520\tablenotemark{e}& 0.119  & 0.045 & 0.75  \\
         &   23, -50 & 1100-1500& 0.351  & 0.272 & 2.48  \\

NGC 7457 &  -42, 35  & 874-974\tablenotemark{a}  & 0.086  & 0.241 & 5.07  \\
         &    0, 0   & 695-945  & 0.234 & 0.073 & 0.913  \\
         &   42, -35 & 674-774  &  0.348 & 0.184 & 3.89  \\

\enddata

\tablecomments{See Table 2 for description of tabulated quantities}

\tablenotetext{a}{Line windows for off-centre pointngs centred on optical
rotation curve from \citet[NGC 936]{knt87}, \citet[NGC 3115]{cap93},
\citet[NGC 4026]{fis97}, and \citet[NGC 7457]{dre83}.}

\tablenotetext{b}{NGC 1023 is possibly merging with a smaller, gas-rich
companion, as the HI diststribution is extensive and complex (Sancisi et al.
1984).  The narrow window at each pointing is based on an estimate of the
local HI velocity from Figures 1 and 3 in Sancisi et al., whereas the wide
window spans the global HI line profile in Figure 2 of the same work.}

\tablenotetext{c}{Insufficient baseline on high-velocity side of line window.}

\tablenotetext{d}{Second-order baseline removed.}

\tablenotetext{e}{Narrow window isolates possible emission.}


\end{deluxetable}
\begin{deluxetable}{lrrcr}
\tablecolumns{5}
\tablewidth{0pc}
\tablecaption{Total H$_2$ Masses and Upper Limits from Our Observations}
\tablehead{
\colhead{Name} & \multicolumn{2}{c}{IRAM 30m} & \colhead{} &
 \colhead{NRAO 12m}\\
\cline{2-3} \\
\colhead{} & \colhead{J=(1-0) \tablenotemark{a}} & 
\colhead{J=(2-1)} & \colhead{} & \colhead{J=(1-0)}
}
\startdata

NGC 936  & \nodata & \nodata           &     & $\pm$1.55$\times$10$^7$    \\
NGC 1023 & \nodata & \nodata           &     & $\pm$4.91$\times$10$^7$      \\
NGC 2787 & \nodata & (6.92$\pm$1.67)$\times$10$^6$   &     &
 (1.78$\pm$0.36)$\times$10$^7$\tablenotemark{b}    \\
NGC 3115 & \nodata & (1.86$\pm$0.52)$\times$10$^6$  &     & \nodata \\
NGC 3384 & \nodata & (3.48$\pm$0.78)$\times$10$^6$  &     & \nodata \\\\

NGC 3412 & (2.22$\pm$0.57)$\times$10$^6$   & (1.38$\pm$0.43)$\times$10$^6$   &     & 
 \nodata \\
NGC 3489 & (6.13$\pm$0.68)$\times$10$^6$   & (4.22$\pm$0.36)$\times$10$^6$   &     & 
 \nodata \\
NGC 3607 & (2.32$\pm$0.10)$\times$10$^8$   & (3.07$\pm$0.08)$\times$10$^8$   &     & 
 (1.03$\pm$0.11)$\times$10$^9$\tablenotemark{b}   \\
NGC 3870 & (1.99$\pm$0.27)$\times$10$^7$   & (2.94$\pm$0.30)$\times$10$^7$   &     & 
 (1.28$\pm$0.31)$\times$10$^8$\tablenotemark{b}   \\                
NGC 3941 & \nodata & \nodata           &      & (6.58$\pm$0.21)$\times$10$^7$   \\\\

NGC 3990 & \nodata & (5.30$\pm$1.11)$\times$10$^6$   &    & \nodata \\
NGC 4026 & \nodata & \nodata           &      & (8.80$\pm$1.84)$\times$10$^7$   \\
NGC 4111 & (1.59$\pm$0.36)$\times$10$^7$   & (2.33$\pm$0.22)$\times$10$^7$   &     &
 \nodata \\
NGC 4143 & $\pm$4.45$\times$10$^6$ & $\pm$5.55$\times$10$^6$             &     &
 \nodata  \\
NGC 4150 & (3.81$\pm$0.12)$\times$10$^7$   & (2.92$\pm$0.18)$\times$10$^7$   &     &
 \nodata \\\\

NGC 4203 & (7.61$\pm$0.85)$\times$10$^6$   & (9.04$\pm$0.70)$\times$10$^6$    &     &
 \nodata \\
NGC 4251 & $\pm$7.87$\times$10$^5$ &  $\pm$8.35$\times$10$^5$             &     & 
 $\pm$6.35$\times$10$^6$\tablenotemark{b}  \\
NGC 4310 & (4.30$\pm$0.17)$\times$10$^7$    & (3.88$\pm$0.16)$\times$10$^7$   &     &
 \nodata  \\
NGC 4346 & $\pm$5.12$\times$10$^6$  &  $\pm$4.46$\times$10$^6$            &     & 
 $\pm$2.78$\times$10$^7$\tablenotemark{b}  \\
NGC 4460 & (1.06$\pm$0.10)$\times$10$^7$   & \nodata  &    &
 (1.32$\pm$0.39)$\times$10$^7$\tablenotemark{b}   \\\\

NGC 4880 & (8.70$\pm$1.12)$\times$10$^6$   & (5.23$\pm$0.58)$\times$10$^6$    &     &
 \nodata \\
NGC 5866 & (4.39$\pm$0.05)$\times$10$^8$\tablenotemark{b}  &
 (5.86$\pm$0.10)$\times$10$^8$    &    & (4.11$\pm$0.35)$\times$10$^8$   \\
NGC 7332 & \nodata  &  \nodata        &     & $\pm$4.16$\times$10$^7$    \\ 
NGC 7457 & \nodata & \nodata          &     &  (3.30$\pm$1.03)$\times$10$^6$   \\
\enddata
\tablenotetext{a} {Columns give $M(H_2)$ and its formal standard 
deviation, or the one-sigma upper limit, all in solar masses.  
Where multiple pointings are available, the total is obtained from successive 
pointings (IRAM 30m, J=2-1; NRAO 12m), or from alterate pointings 
(IRAM 30m, J=1-0).  Standard deviations are added in quadrature.}  

\tablenotetext{b} {Used in Table 6.}
\end{deluxetable}

\begin{deluxetable}{ccccc}
\tabletypesize{\scriptsize}
\tablecolumns{5}
\tablewidth{0pc}
\tablecaption{Comparison with Published Measurements}
\tablehead{
\colhead{Name}  &  \multicolumn{2}{c}{Other 
Studies}  &  \multicolumn{2}{c}{Present Study} \\
\colhead{}      &  \colhead{$M($H$_2$)}    &  \colhead{FWHM}  & 
\colhead{$M($H$_2$)}  &
\colhead{FWHM\tablenotemark{a}} \\
\colhead{}  &  \colhead{$M_{\sun}$}    &  \colhead{$\arcsec$} 
&\colhead{$M_{\sun}$}    &
\colhead{$\arcsec$}  }
\startdata


NGC 1023  &  $<$1.69$\times$10$^6$\tablenotemark{b}   &   15   & 
$<$7.71$\times$10$^6$\tablenotemark{c,d}   &  55   \\
NGC 3115  &   $<$4.35$\times$10$^6$\tablenotemark{b}    &   15   & 
1.864$\times$10$^6$\tablenotemark{f}   &  10   \\
           &   \nodata     &  \nodata    & 
$<$3.37$\times$10$^6$\tablenotemark{g}   &  21   \\
           &   \nodata     &  \nodata   &    $<$2.33$\times$10$^7$   &  55   \\
NGC 4251  & $<$1.1$\times$10$^7$\tablenotemark{h}  &         21    & 
$<$2.36$\times$10$^6$   &  21    \\
           &   \nodata     &  \nodata    &    $<$1.94$\times$10$^7$   &  55 
   \\
NGC 4310  &   6.30$\times$10$^7$\tablenotemark{j}    &    21  & 
2.702$\times$10$^7$   &  21   \\
           &   \nodata     &  \nodata &    $<$5.96$\times$10$^7$   &  55    \\
NGC 5866  &   8.34$\times$10$^7$\tablenotemark{b}    &    15  & 
2.17$\times$10$^8$    &  21    \\
           &   1.02$\times$10$^8$\tablenotemark{m}    &    21 
&     2.17$\times$10$^8$    &  21    \\
           &   5.09$\times$10$^8$\tablenotemark{n}   &    45  & 
3.58$\times$10$^8$    &  55 \\
           &   4.37$\times$10$^8$\tablenotemark{o}    &    3 
$\times$45  &   4.38$\times$10$^8$  &  5 $\times$ 21  \\
NGC 7457  &   2.64$\times$10$^7$\tablenotemark{b}    &    15  & 
3.30$\times$10$^6$    &    55    \\
\enddata
\tablecomments{ Values from center pointings are compared unless otherwise
indicated.}
\tablenotetext{a}{"10" indicates 30m J=2-1, "21" indicates 30m J=1-0, "55" 
indicates 12m J=1-0.}
\tablenotetext{b}{\citet{tan94}, Nobeyama 45m telescope.}
\tablenotetext{c}{From wide line window at (0,0).}
\tablenotetext{d}{Upper limits from the present study are 3-sigma.}
\tablenotetext{f}{Using J=2-1 transition at (0,0).}
\tablenotetext{g}{From J=1-0 transition at (0,0).}
\tablenotetext{h}{\citet{ger94}, IRAM 30m telescope}
\tablenotetext{j}{\citet{ger94}.  Possibly the sum of as many as 6 pointings.}
\tablenotetext{k}{From the sum of (-17,14), (0,0), and (17,-14)}
\tablenotetext{m}{WH, IRAM 30m telescope.}
\tablenotetext{n}{\citet{thr89}, FCRAO 13.7m telescope.}
\tablenotetext{o}{\citet{yng95}, FCRAO 13.7m telescope.}

\end{deluxetable}

\begin{deluxetable}{lcccc}
\tabletypesize{\footnotesize}
\tablecolumns{5}
\tablewidth{0pc}
\tablecaption{Observed and Predicted Cool ISM Masses}
\tablehead{
\colhead{Name} & \colhead{M$_r$} & \colhead{M$(H_2)$} &
\colhead{M$_{obs}$} & \colhead{HI} \\
\colhead{} & \colhead{($M_{\sun}$)} & \colhead{($M_{\sun}$)} &
\colhead{($M_{\sun}$)} & \colhead{ref} \\
}
\startdata


NGC 404   & 4.48$\times$10$^8$    & \nodata 
 & 1.01$\times$10$^8$ \tablenotemark{a}    & 1 \\
NGC 936   & 8.28$\times$10$^9$    & 
$<$1.55$\times$10$^7$   & $<$7.33$\times$10$^8$\tablenotemark{b}  & 1 \\
NGC 1023  & 1.86$\times$10$^{10}$ &   $<$4.91$\times$10$^7$
\tablenotemark{c} &    $<$2.49$\times$10$^9$    & 2 \\
NGC 2787  & 6.97$\times$10$^9$    & 
1.78$\times$10$^7$     &    1.10$\times$10$^9$   & 3  \\
NGC 3115  & 1.04$\times$10$^{10}$ & 1.86$\times$10$^6$\tablenotemark{e} & 
 $<$1.86$\times$10$^7$  & 4 \\\\

NGC 3384  & 5.97$\times$10$^9$    & 3.48$\times$10$^6$\tablenotemark{e} &
  $<$2.22$\times$10$^7$  & 5 \\
NGC 3412  & 3.47$\times$10$^9$    & 2.22$\times$10$^6$  & 
  $<$1.82$\times$10$^7$  & 5 \\
NGC 3489  & 2.58$\times$10$^9$    & 6.13$\times$10$^6$  & 
  1.67$\times$10$^7$  & 5 \\
NGC 3607  & 3.47$\times$10$^{10}$ &
1.03$\times$10$^9$           &  $<$1.50$\times$10$^9$  & 6 \\
NGC 3870  & 2.15$\times$10$^9$    &
1.28$\times$10$^8$           &  $<$1.86$\times$10$^9$  & 7 \\\\

NGC 3941  & 2.05$\times$10$^{10}$ & 
6.58$\times$10$^7$           &     1.94$\times$10$^9$  & 8 \\
NGC 3990  & 2.17$\times$10$^9$    & 5.30$\times$10$^6$\tablenotemark{e} & 
 \nodata\tablenotemark{f}  & \nodata \\
NGC 4026  & 1.21$\times$10$^{10}$ & 
8.80$\times$10$^7$\tablenotemark{g} & $<$2.15$\times$10$^8$  & 9 \\
NGC 4111  & 1.20$\times$10$^{10}$ & 1.59$\times$10$^7$  & 
 1.13$\times$10$^9$      & 10 \\
NGC 4143  & 9.12$\times$10$^9$    & 
$<$4.45$\times$10$^6$  &
$<$1.00$\times$10$^9$   & 8 \\\\

NGC 4150  & 1.87$\times$10$^9$    & 
3.81$\times$10$^7$  &
$<$6.27$\times$10$^7$   & 11 \\
NGC 4203  & 3.88$\times$10$^9$    & 
7.61$\times$10$^6$  &    7.43$\times$10$^8$
  & 12 \\
NGC 4251  & 4.58$\times$10$^9$    & 
  $<$6.35$\times$10$^6$       & 
$<$3.89$\times$10$^7$   & 13 \\
NGC 4310  & 1.00$\times$10$^9$    &
4.30$\times$10$^7$  &   9.59$\times$10$^7$
  & 14 \\
NGC 4346  & 7.11$\times$10$^9$    &
 $<$2.78$\times$10$^7$       &
$<$5.17$\times$10$^8$   & 15 \\\\

NGC 4460  & 
1.50$\times$10$^9$    &  1.32$\times$10$^7$ 
& \nodata\tablenotemark{f}  &  \nodata \\
NGC 4880  & 5.14$\times$10$^9$    &  8.70$\times$10$^6$ 
&  $<$2.69$\times$10$^7$   & 9 \\
NGC 5866  & 1.98$\times$10$^{10}$ &  4.39$\times$10$^8$ 
&  $<$1.32$\times$10$^9$   & 16 \\
NGC 7013  & 1.92$\times$10$^{10}$ 
& \nodata                     & 
1.57$\times$10$^9$\tablenotemark{h}  & 17 \\
NGC 7077  & 
8.53$\times$10$^8$    &  \nodata 
& 1.22$\times$10$^8$\tablenotemark{i}  & 18 \\\\

NGC 7332  & 1.02$\times$10$^{10}$ &  
$<$4.16$\times$10$^7$        & $<$1.98$\times$10$^8$   & 19 \\
NGC 7457  & 5.43$\times$10$^9$    &  
3.30$\times$10$^6$        & $<$9.28$\times$10$^7$   & 14 \\

\enddata
\tiny
\tablecomments{Columns contain galaxy name, total mass returned by evolving 
stars over last 10 Gyr,
estimated as described in the text, H$_2$ mass or 1-$\sigma$ limit from 
present observations,
observed cool ISM mass including He, reference for HI observations.}

\tablenotetext{a}{CO observations from \citet{sag89}.}

\tablenotetext{b}{Upper limits to M$_{obs}$ are the sum of $M($H$_2)$, or 
its 3-$\sigma$ limit, and the
3-$\sigma$ limit on HI mass from the indicated reference.  H$_2$ has been 
detected in every galaxy known to contain HI, except for NGC 1023.}

\tablenotetext{c}{From wide line windows at all offsets}

\tablenotetext{d}{From CO measurement plus 3-sigma HI limit}

\tablenotetext{e}{From J=2-1 transition}

\tablenotetext{f}{No HI observation.}

\tablenotetext{g}{Detection at offset (2,-55).}

\tablenotetext{h}{CO observations from \citet{yng95}.}

\tablenotetext{i}{CO observations from \citet{sag92}.}

\tablerefs{         (1) \citep{war86};  (2) \citet{san84};
(3) \citet{sho87};  (4) \citet{ric87};  (5) \citet{gio83};  (6) \citet{bie79};
(7) \citet{bot73};  (8) \citet{huc82};  (9) \citet{rob91};  (10)
\citet{bur85};
(11) \citet{kna79}; (12) \citet{vdr88}; (13) \citet{bie77}; (14)
\citet{cha87};
(15) \citet{app82}; (16) \citet{kng82}; (17) \citet{kna84}; (18) \citet{lew87};
(19) \citet{bur87}.  }

\end{deluxetable}

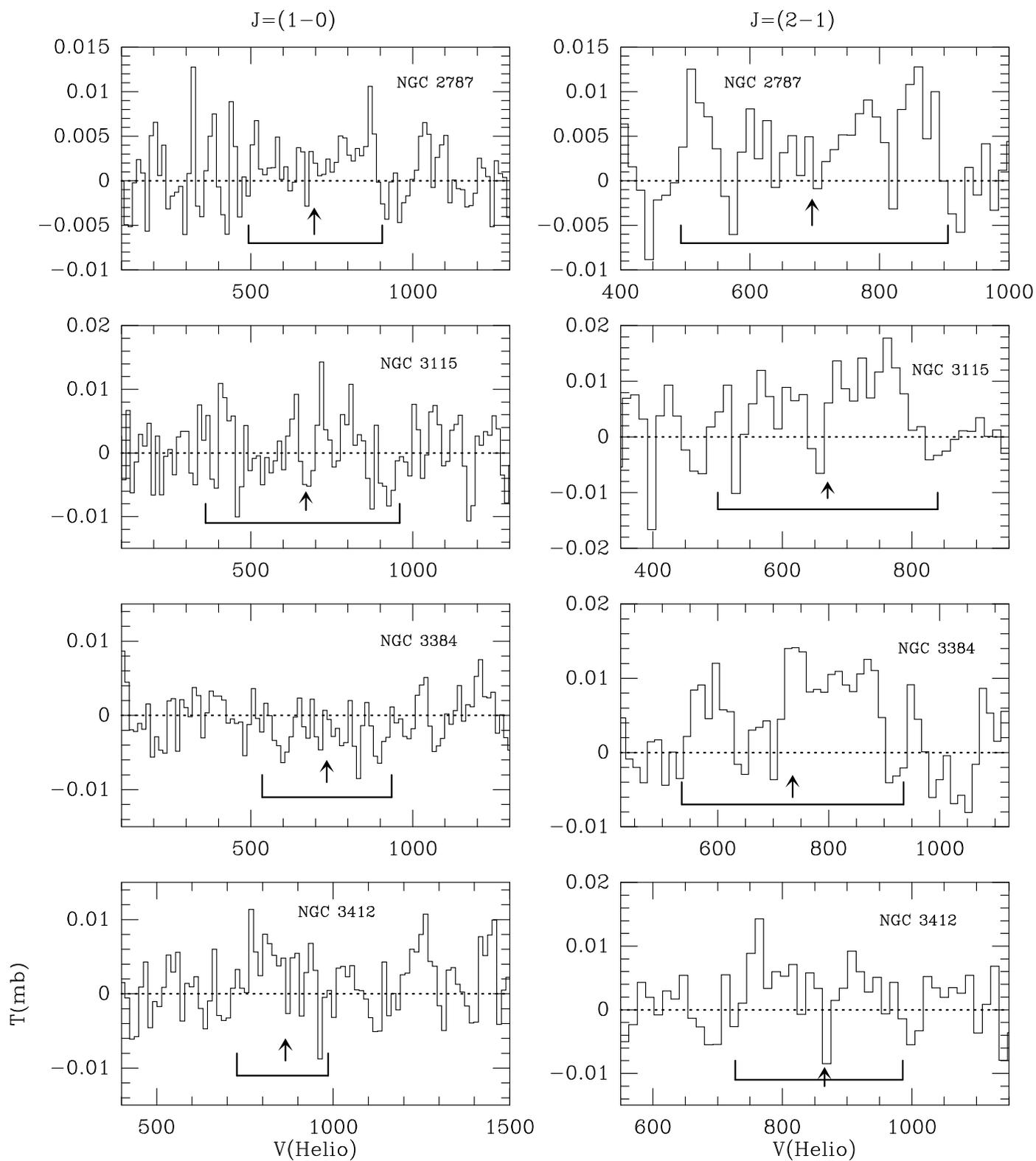
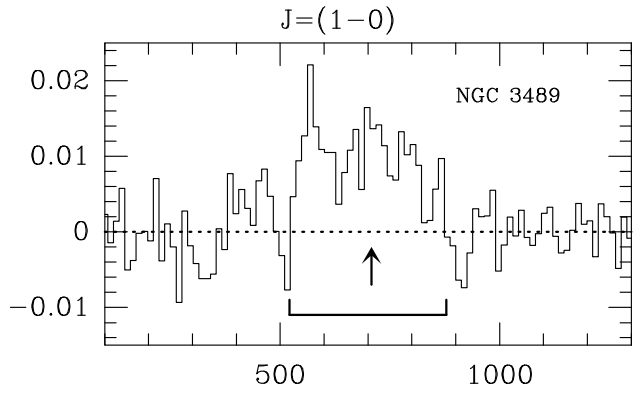
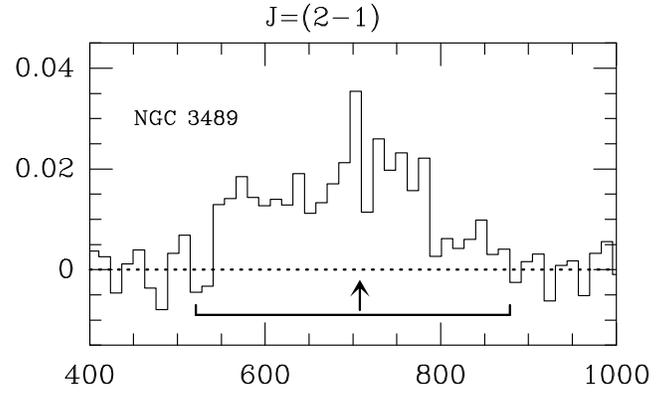
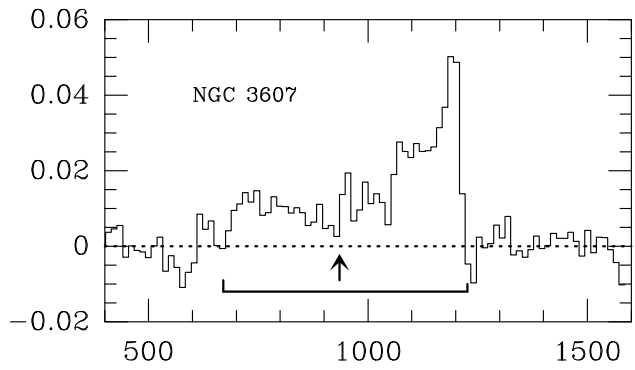
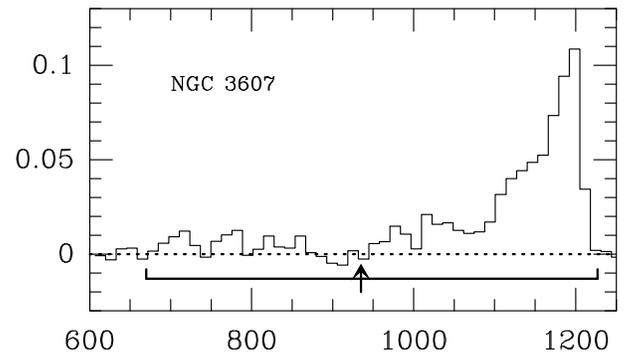
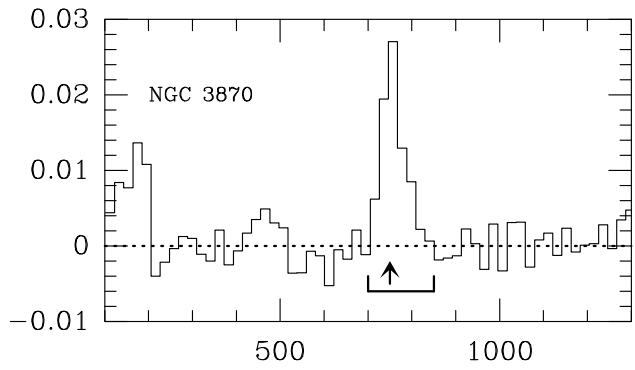
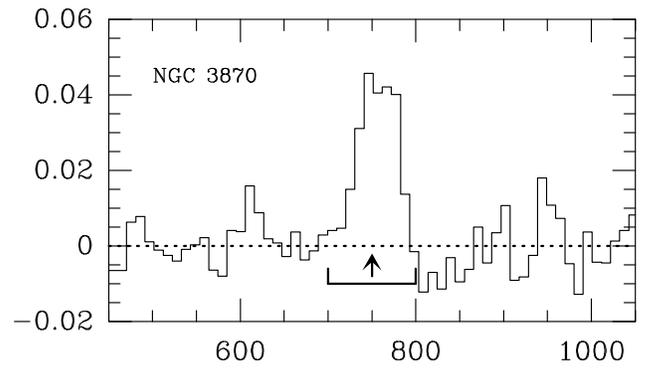
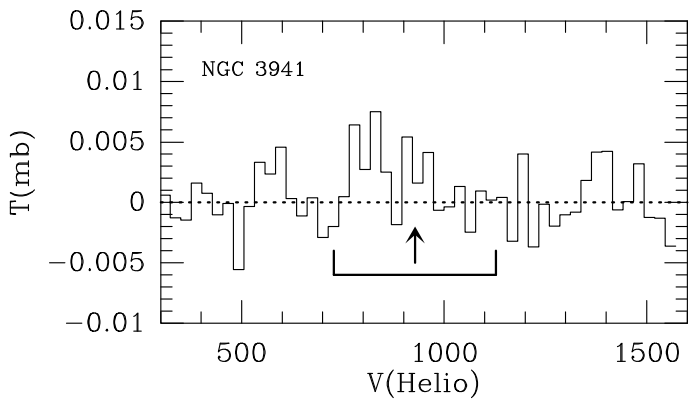
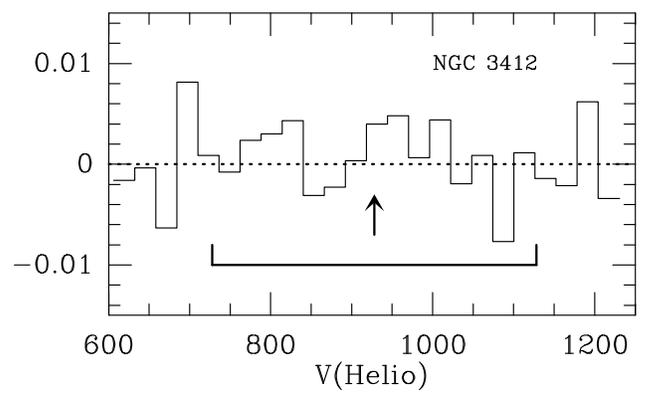
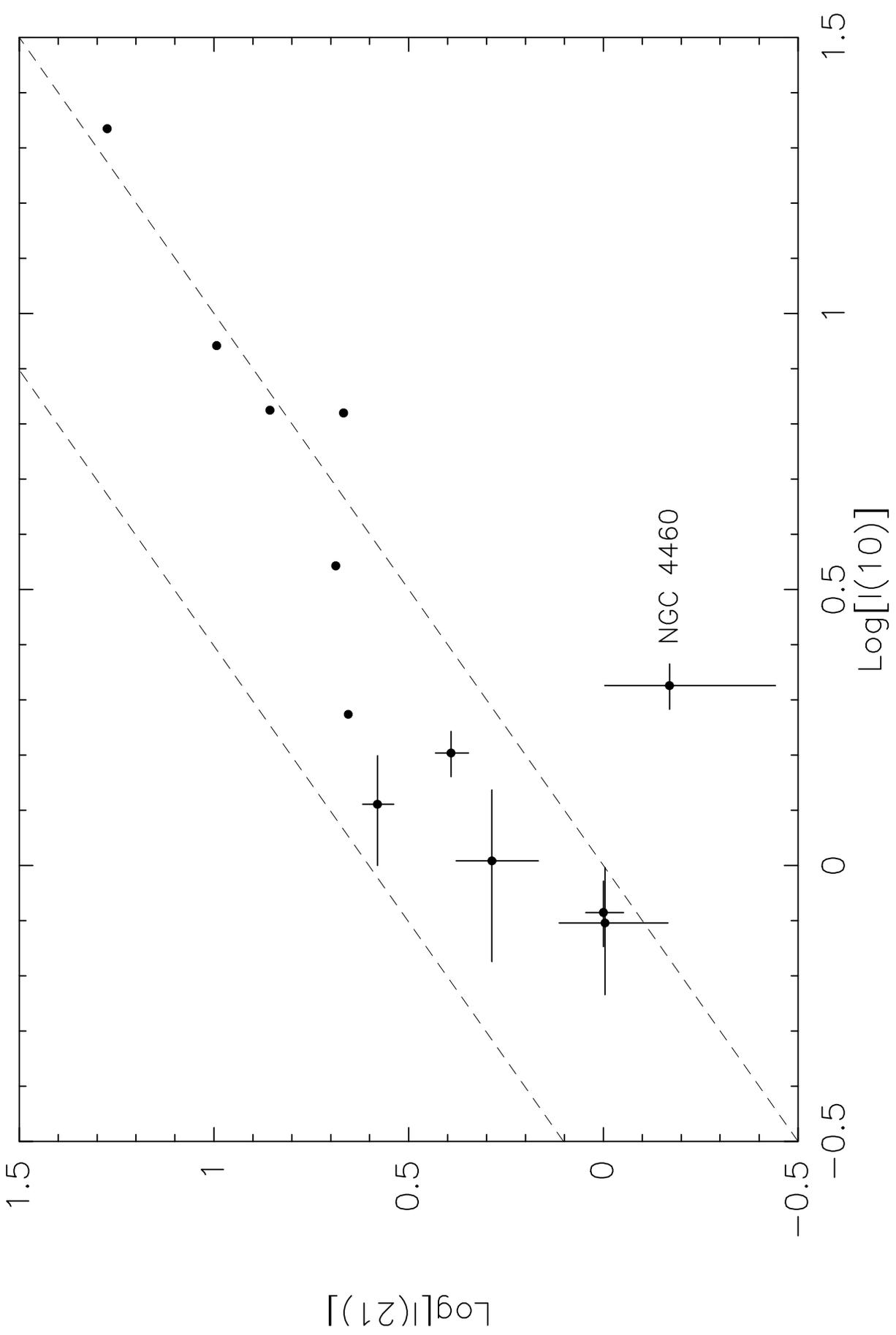
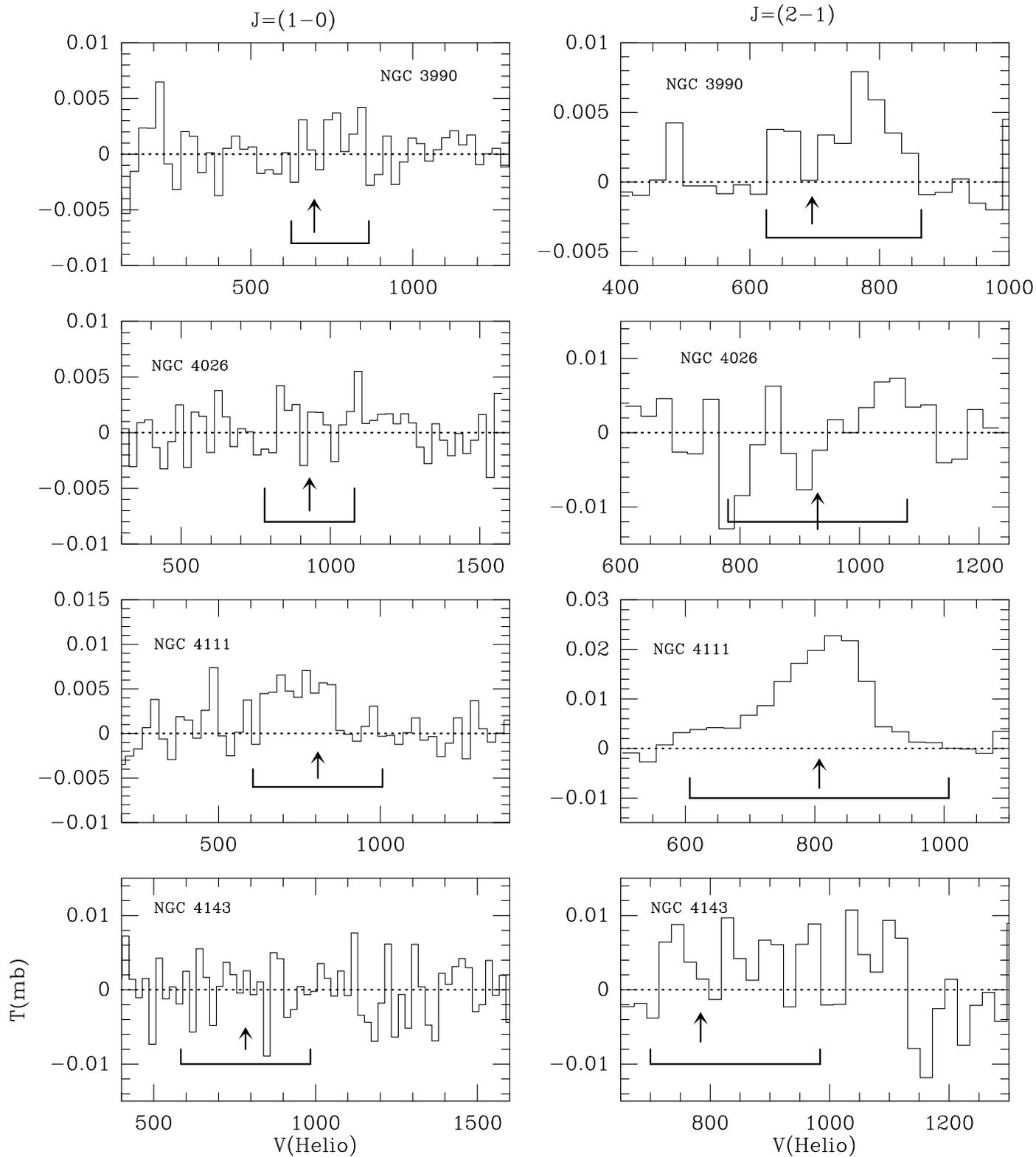
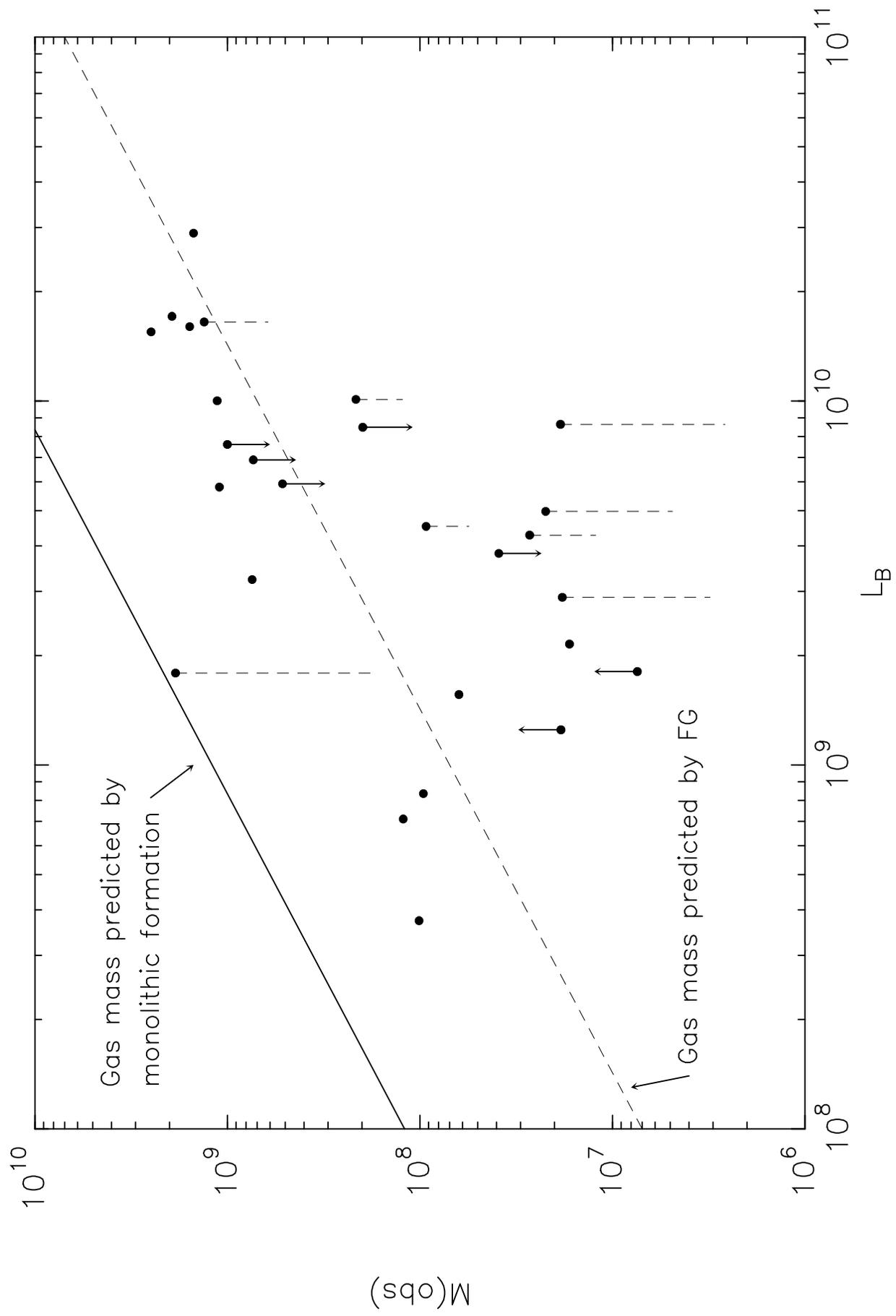
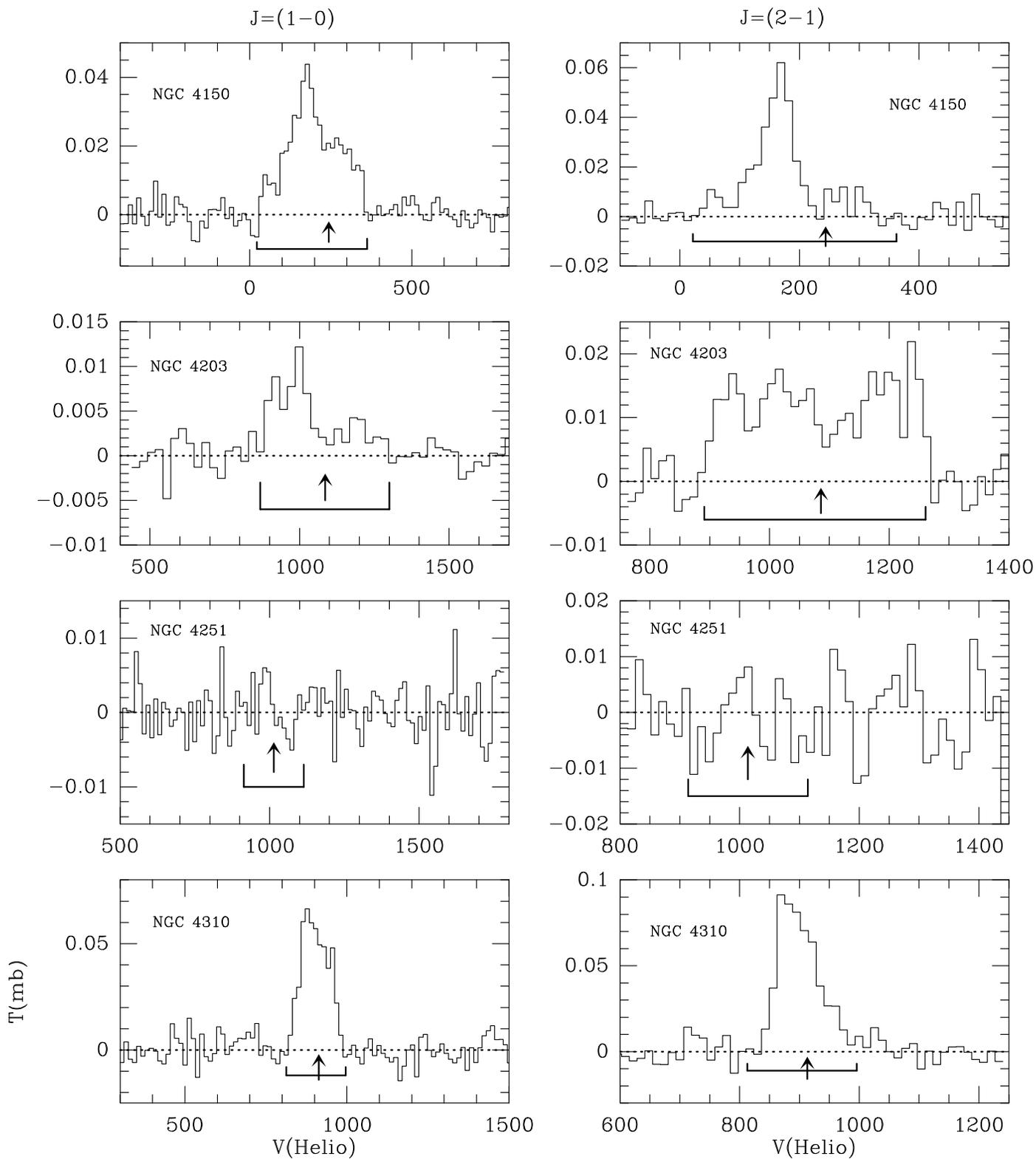
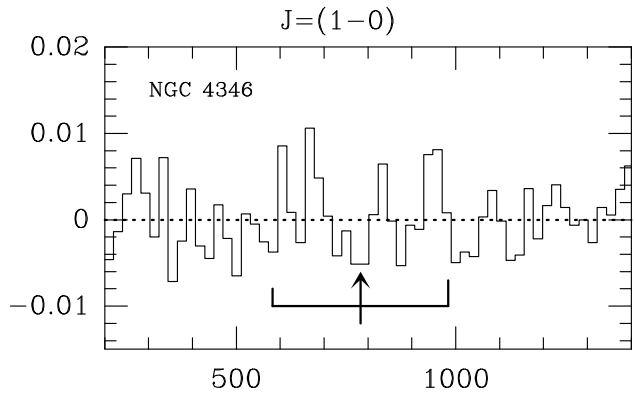
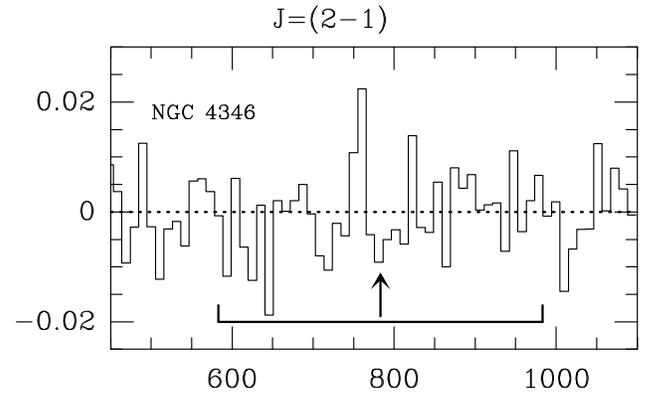
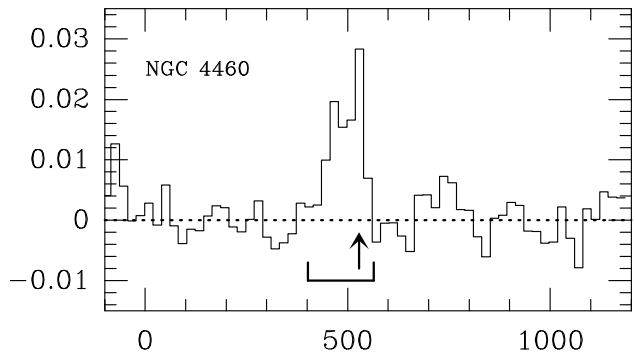
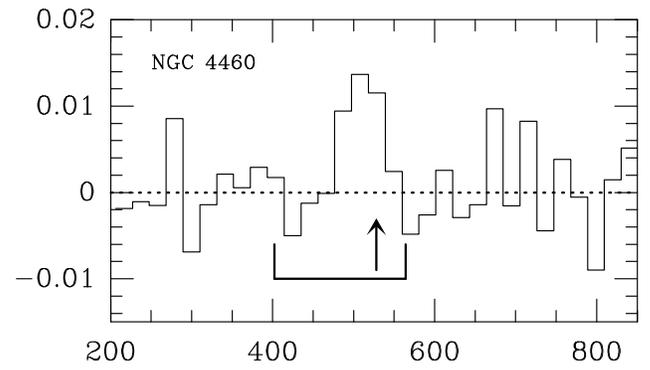
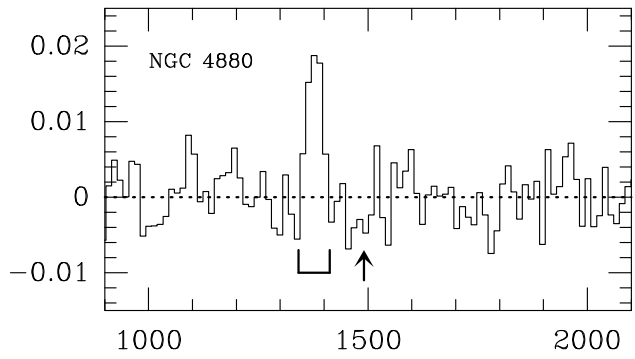
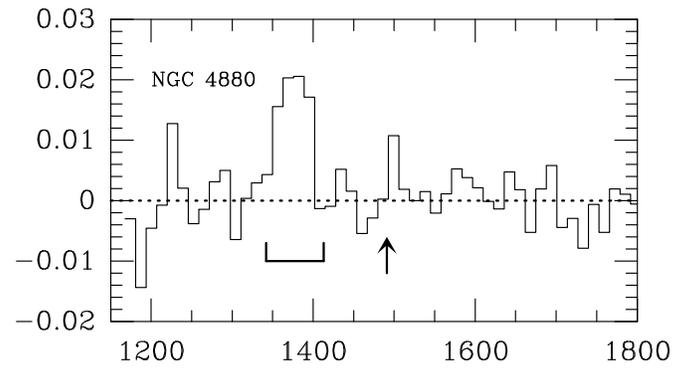
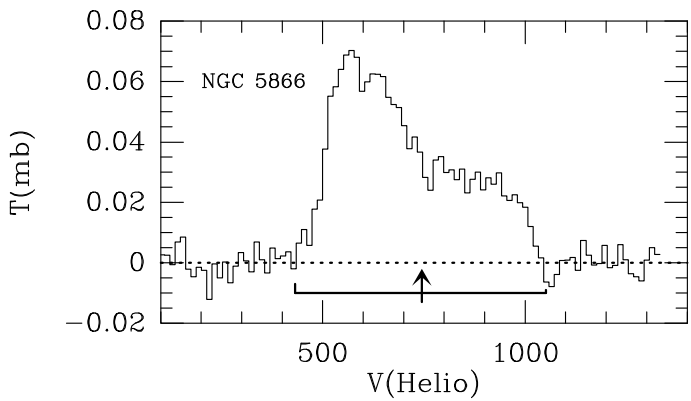
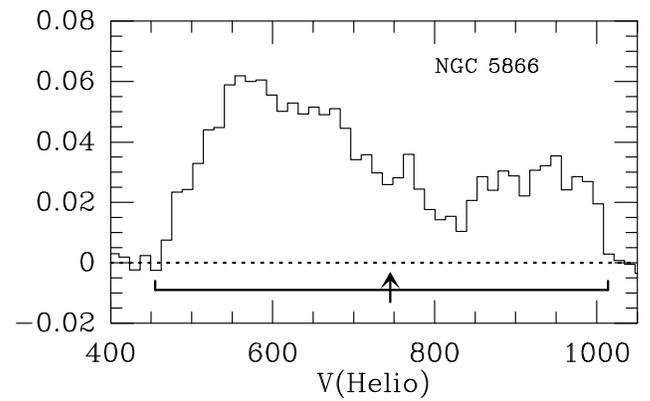
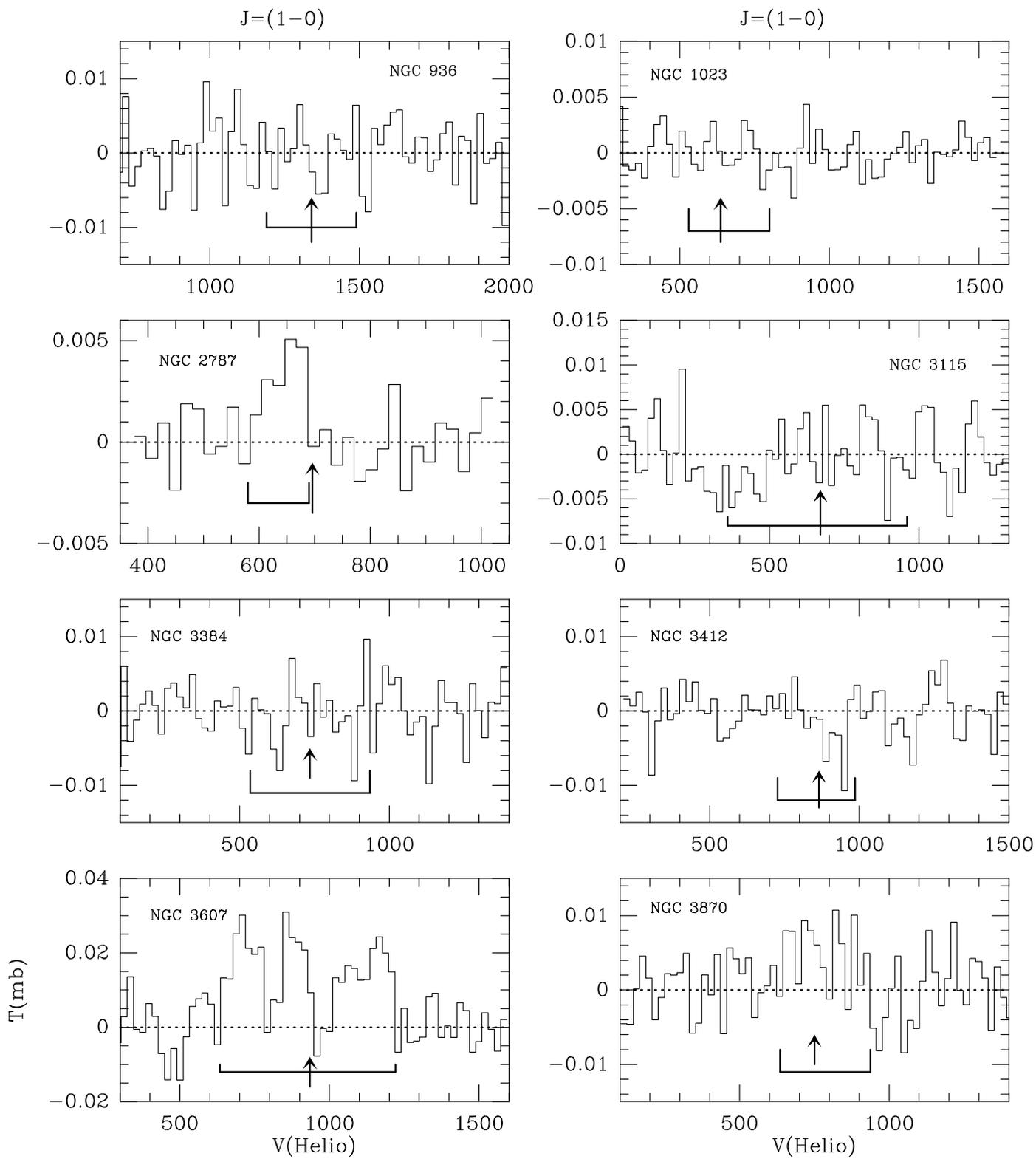
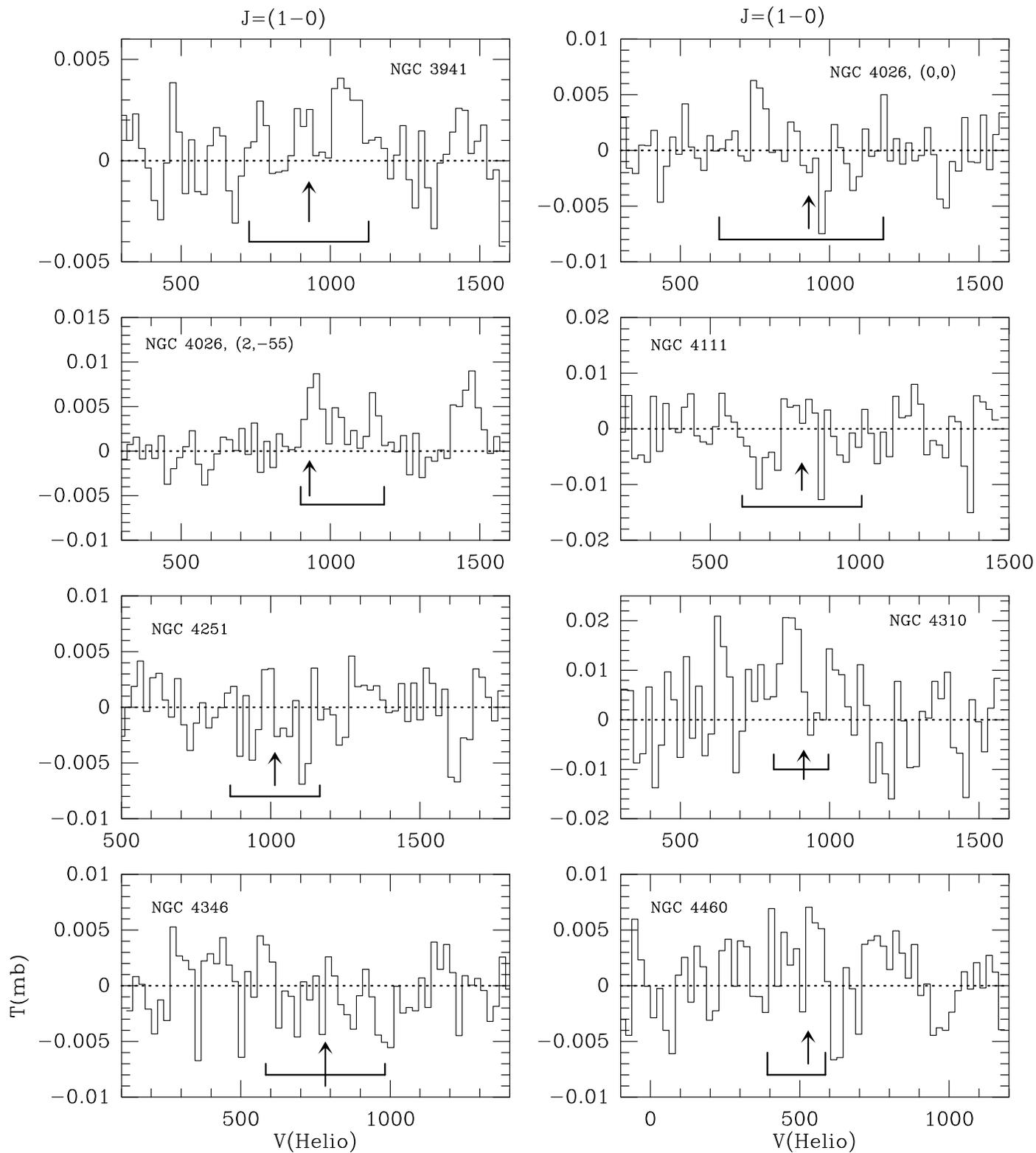
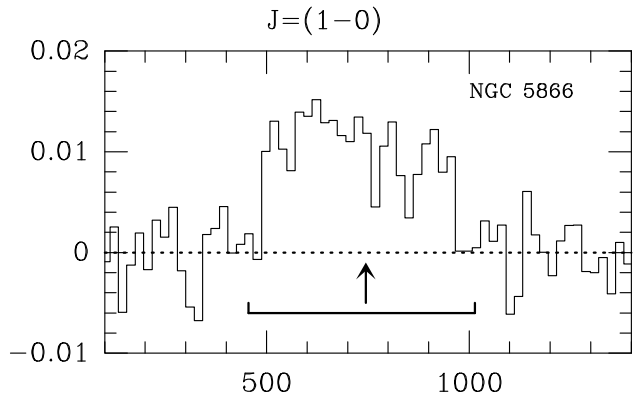
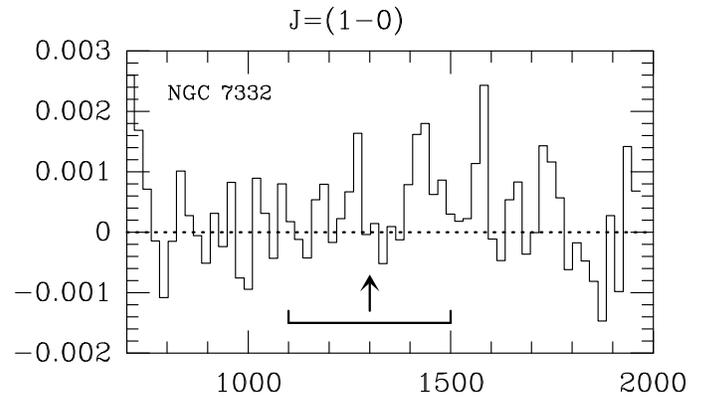
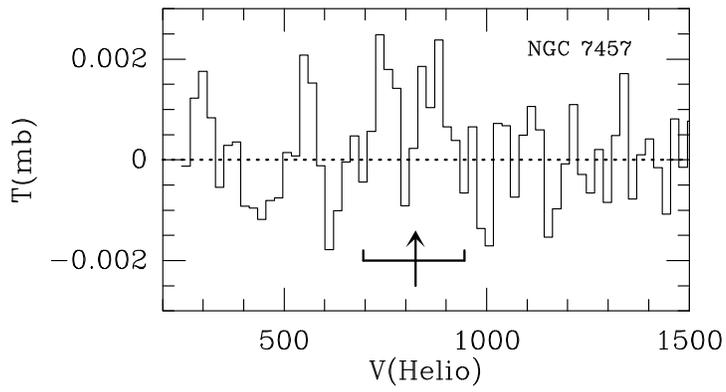
\end{document}